\documentclass[longauth]{aaEC}

\usepackage{ISTL_macros}

\usepackage{bm}

\usepackage[english]{babel}
\usepackage{natbib}
\bibpunct{(}{)}{;}{a}{}{,} 
\usepackage{amsmath}
\usepackage{graphicx}
\usepackage{txfonts}
\usepackage[colorlinks=true, allcolors=blue]{hyperref}
\usepackage{enumitem}
\usepackage{verbatim}
\usepackage{url}
\usepackage{color, colortbl}
\usepackage{amsfonts}
\usepackage{amssymb}
\usepackage{bm}
\usepackage{epsfig}
\usepackage{mathtools}
\usepackage[normalem]{ulem}
\usepackage{multirow}
\usepackage{tabularx}
\usepackage{booktabs}
\usepackage[nameinlink,capitalise]{cleveref}
\usepackage{euclid}
\usepackage{adjustbox}
\usepackage{placeins}
\usepackage{xurl}
\usepackage{lastpage}

\begin{document}
\title{\Euclid preparation}
\subtitle{Cosmology Likelihood for Observables in Euclid (CLOE). 6: Impact of systematic uncertainties on the cosmological analysis}    

\newcommand{\orcid}[1]{} 
\author{Euclid Collaboration: L.~Blot\orcid{0000-0002-9622-7167}\thanks{\email{linda.blot@ipmu.jp}}\inst{\ref{aff1},\ref{aff2}}
\and K.~Tanidis\orcid{0000-0001-9843-5130}\inst{\ref{aff3}}
\and G.~Ca\~nas-Herrera\orcid{0000-0003-2796-2149}\inst{\ref{aff4},\ref{aff5}}
\and P.~Carrilho\orcid{0000-0003-1339-0194}\inst{\ref{aff6},\ref{aff7}}
\and M.~Bonici\orcid{0000-0002-8430-126X}\inst{\ref{aff8},\ref{aff9}}
\and S.~Camera\orcid{0000-0003-3399-3574}\inst{\ref{aff10},\ref{aff11},\ref{aff12}}
\and V.~F.~Cardone\inst{\ref{aff13},\ref{aff14}}
\and S.~Casas\orcid{0000-0002-4751-5138}\inst{\ref{aff15},\ref{aff16}}
\and S.~Davini\orcid{0000-0003-3269-1718}\inst{\ref{aff17}}
\and S.~Di~Domizio\orcid{0000-0003-2863-5895}\inst{\ref{aff18},\ref{aff17}}
\and S.~Farrens\orcid{0000-0002-9594-9387}\inst{\ref{aff19}}
\and L.~W.~K.~Goh\orcid{0000-0002-0104-8132}\inst{\ref{aff19}}
\and S.~Gouyou~Beauchamps\inst{\ref{aff20},\ref{aff21}}
\and S.~Ili\'c\orcid{0000-0003-4285-9086}\inst{\ref{aff22},\ref{aff23}}
\and S.~Joudaki\orcid{0000-0001-8820-673X}\inst{\ref{aff24}}
\and F.~Keil\orcid{0000-0002-8108-1679}\inst{\ref{aff23}}
\and A.~M.~C.~Le~Brun\orcid{0000-0002-0936-4594}\inst{\ref{aff2}}
\and M.~Martinelli\orcid{0000-0002-6943-7732}\inst{\ref{aff13},\ref{aff14}}
\and C.~Moretti\orcid{0000-0003-3314-8936}\inst{\ref{aff25},\ref{aff26},\ref{aff27},\ref{aff28}}
\and V.~Pettorino\orcid{0000-0002-4203-9320}\inst{\ref{aff4}}
\and A.~Pezzotta\orcid{0000-0003-0726-2268}\inst{\ref{aff29}}
\and Z.~Sakr\orcid{0000-0002-4823-3757}\inst{\ref{aff30},\ref{aff23},\ref{aff31}}
\and A.~G.~S\'anchez\orcid{0000-0003-1198-831X}\inst{\ref{aff32}}
\and D.~Sciotti\orcid{0009-0008-4519-2620}\inst{\ref{aff13},\ref{aff14}}
\and I.~Tutusaus\orcid{0000-0002-3199-0399}\inst{\ref{aff21},\ref{aff20},\ref{aff23}}
\and V.~Ajani\orcid{0000-0001-9442-2527}\inst{\ref{aff19},\ref{aff33},\ref{aff34}}
\and M.~Crocce\orcid{0000-0002-9745-6228}\inst{\ref{aff21},\ref{aff20}}
\and A.~Fumagalli\orcid{0009-0004-0300-2535}\inst{\ref{aff25}}
\and C.~Giocoli\orcid{0000-0002-9590-7961}\inst{\ref{aff35},\ref{aff36}}
\and L.~Legrand\orcid{0000-0003-0610-5252}\inst{\ref{aff37},\ref{aff38}}
\and M.~Lembo\orcid{0000-0002-5271-5070}\inst{\ref{aff39},\ref{aff40},\ref{aff41}}
\and G.~F.~Lesci\orcid{0000-0002-4607-2830}\inst{\ref{aff42},\ref{aff35}}
\and D.~Navarro~Girones\orcid{0000-0003-0507-372X}\inst{\ref{aff5}}
\and A.~Nouri-Zonoz\orcid{0009-0006-6164-8670}\inst{\ref{aff43}}
\and S.~Pamuk\orcid{0009-0004-0852-8624}\inst{\ref{aff44}}
\and A.~Pourtsidou\orcid{0000-0001-9110-5550}\inst{\ref{aff7},\ref{aff45}}
\and M.~Tsedrik\orcid{0000-0002-0020-5343}\inst{\ref{aff7},\ref{aff45}}
\and J.~Bel\inst{\ref{aff46}}
\and C.~Carbone\orcid{0000-0003-0125-3563}\inst{\ref{aff9}}
\and C.~A.~J.~Duncan\orcid{0009-0003-3573-0791}\inst{\ref{aff7}}
\and M.~Kilbinger\orcid{0000-0001-9513-7138}\inst{\ref{aff19}}
\and D.~Sapone\orcid{0000-0001-7089-4503}\inst{\ref{aff47}}
\and E.~Sellentin\inst{\ref{aff48},\ref{aff5}}
\and P.~L.~Taylor\orcid{0000-0001-6999-4718}\inst{\ref{aff49},\ref{aff50}}
\and L.~Amendola\orcid{0000-0002-0835-233X}\inst{\ref{aff30}}
\and S.~Andreon\orcid{0000-0002-2041-8784}\inst{\ref{aff29}}
\and N.~Auricchio\orcid{0000-0003-4444-8651}\inst{\ref{aff35}}
\and C.~Baccigalupi\orcid{0000-0002-8211-1630}\inst{\ref{aff26},\ref{aff25},\ref{aff27},\ref{aff28}}
\and M.~Baldi\orcid{0000-0003-4145-1943}\inst{\ref{aff51},\ref{aff35},\ref{aff36}}
\and S.~Bardelli\orcid{0000-0002-8900-0298}\inst{\ref{aff35}}
\and P.~Battaglia\orcid{0000-0002-7337-5909}\inst{\ref{aff35}}
\and A.~Biviano\orcid{0000-0002-0857-0732}\inst{\ref{aff25},\ref{aff26}}
\and E.~Branchini\orcid{0000-0002-0808-6908}\inst{\ref{aff18},\ref{aff17},\ref{aff29}}
\and M.~Brescia\orcid{0000-0001-9506-5680}\inst{\ref{aff52},\ref{aff53}}
\and V.~Capobianco\orcid{0000-0002-3309-7692}\inst{\ref{aff12}}
\and J.~Carretero\orcid{0000-0002-3130-0204}\inst{\ref{aff24},\ref{aff54}}
\and M.~Castellano\orcid{0000-0001-9875-8263}\inst{\ref{aff13}}
\and G.~Castignani\orcid{0000-0001-6831-0687}\inst{\ref{aff35}}
\and S.~Cavuoti\orcid{0000-0002-3787-4196}\inst{\ref{aff53},\ref{aff55}}
\and K.~C.~Chambers\orcid{0000-0001-6965-7789}\inst{\ref{aff56}}
\and A.~Cimatti\inst{\ref{aff57}}
\and C.~Colodro-Conde\inst{\ref{aff58}}
\and G.~Congedo\orcid{0000-0003-2508-0046}\inst{\ref{aff7}}
\and C.~J.~Conselice\orcid{0000-0003-1949-7638}\inst{\ref{aff59}}
\and L.~Conversi\orcid{0000-0002-6710-8476}\inst{\ref{aff60},\ref{aff61}}
\and Y.~Copin\orcid{0000-0002-5317-7518}\inst{\ref{aff62}}
\and F.~Courbin\orcid{0000-0003-0758-6510}\inst{\ref{aff63},\ref{aff64},\ref{aff65}}
\and H.~M.~Courtois\orcid{0000-0003-0509-1776}\inst{\ref{aff66}}
\and M.~Cropper\orcid{0000-0003-4571-9468}\inst{\ref{aff67}}
\and A.~Da~Silva\orcid{0000-0002-6385-1609}\inst{\ref{aff68},\ref{aff69}}
\and H.~Degaudenzi\orcid{0000-0002-5887-6799}\inst{\ref{aff70}}
\and S.~de~la~Torre\inst{\ref{aff71}}
\and G.~De~Lucia\orcid{0000-0002-6220-9104}\inst{\ref{aff25}}
\and H.~Dole\orcid{0000-0002-9767-3839}\inst{\ref{aff72}}
\and M.~Douspis\orcid{0000-0003-4203-3954}\inst{\ref{aff72}}
\and F.~Dubath\orcid{0000-0002-6533-2810}\inst{\ref{aff70}}
\and X.~Dupac\inst{\ref{aff61}}
\and S.~Dusini\orcid{0000-0002-1128-0664}\inst{\ref{aff73}}
\and S.~Escoffier\orcid{0000-0002-2847-7498}\inst{\ref{aff74}}
\and M.~Farina\orcid{0000-0002-3089-7846}\inst{\ref{aff75}}
\and F.~Faustini\orcid{0000-0001-6274-5145}\inst{\ref{aff13},\ref{aff76}}
\and S.~Ferriol\inst{\ref{aff62}}
\and F.~Finelli\orcid{0000-0002-6694-3269}\inst{\ref{aff35},\ref{aff77}}
\and M.~Frailis\orcid{0000-0002-7400-2135}\inst{\ref{aff25}}
\and E.~Franceschi\orcid{0000-0002-0585-6591}\inst{\ref{aff35}}
\and M.~Fumana\orcid{0000-0001-6787-5950}\inst{\ref{aff9}}
\and S.~Galeotta\orcid{0000-0002-3748-5115}\inst{\ref{aff25}}
\and K.~George\orcid{0000-0002-1734-8455}\inst{\ref{aff78}}
\and W.~Gillard\orcid{0000-0003-4744-9748}\inst{\ref{aff74}}
\and B.~Gillis\orcid{0000-0002-4478-1270}\inst{\ref{aff7}}
\and J.~Gracia-Carpio\inst{\ref{aff32}}
\and A.~Grazian\orcid{0000-0002-5688-0663}\inst{\ref{aff79}}
\and F.~Grupp\inst{\ref{aff32},\ref{aff80}}
\and S.~V.~H.~Haugan\orcid{0000-0001-9648-7260}\inst{\ref{aff81}}
\and H.~Hoekstra\orcid{0000-0002-0641-3231}\inst{\ref{aff5}}
\and W.~Holmes\inst{\ref{aff82}}
\and I.~M.~Hook\orcid{0000-0002-2960-978X}\inst{\ref{aff83}}
\and F.~Hormuth\inst{\ref{aff84}}
\and A.~Hornstrup\orcid{0000-0002-3363-0936}\inst{\ref{aff85},\ref{aff86}}
\and K.~Jahnke\orcid{0000-0003-3804-2137}\inst{\ref{aff87}}
\and M.~Jhabvala\inst{\ref{aff88}}
\and B.~Joachimi\orcid{0000-0001-7494-1303}\inst{\ref{aff89}}
\and E.~Keih\"anen\orcid{0000-0003-1804-7715}\inst{\ref{aff90}}
\and S.~Kermiche\orcid{0000-0002-0302-5735}\inst{\ref{aff74}}
\and B.~Kubik\orcid{0009-0006-5823-4880}\inst{\ref{aff62}}
\and M.~K\"ummel\orcid{0000-0003-2791-2117}\inst{\ref{aff80}}
\and M.~Kunz\orcid{0000-0002-3052-7394}\inst{\ref{aff43}}
\and H.~Kurki-Suonio\orcid{0000-0002-4618-3063}\inst{\ref{aff91},\ref{aff92}}
\and O.~Lahav\orcid{0000-0002-1134-9035}\inst{\ref{aff89}}
\and S.~Ligori\orcid{0000-0003-4172-4606}\inst{\ref{aff12}}
\and P.~B.~Lilje\orcid{0000-0003-4324-7794}\inst{\ref{aff81}}
\and V.~Lindholm\orcid{0000-0003-2317-5471}\inst{\ref{aff91},\ref{aff92}}
\and I.~Lloro\orcid{0000-0001-5966-1434}\inst{\ref{aff93}}
\and G.~Mainetti\orcid{0000-0003-2384-2377}\inst{\ref{aff94}}
\and D.~Maino\inst{\ref{aff95},\ref{aff9},\ref{aff96}}
\and E.~Maiorano\orcid{0000-0003-2593-4355}\inst{\ref{aff35}}
\and O.~Mansutti\orcid{0000-0001-5758-4658}\inst{\ref{aff25}}
\and O.~Marggraf\orcid{0000-0001-7242-3852}\inst{\ref{aff97}}
\and K.~Markovic\orcid{0000-0001-6764-073X}\inst{\ref{aff82}}
\and N.~Martinet\orcid{0000-0003-2786-7790}\inst{\ref{aff71}}
\and F.~Marulli\orcid{0000-0002-8850-0303}\inst{\ref{aff42},\ref{aff35},\ref{aff36}}
\and R.~J.~Massey\orcid{0000-0002-6085-3780}\inst{\ref{aff98}}
\and E.~Medinaceli\orcid{0000-0002-4040-7783}\inst{\ref{aff35}}
\and S.~Mei\orcid{0000-0002-2849-559X}\inst{\ref{aff99},\ref{aff100}}
\and Y.~Mellier\inst{\ref{aff101},\ref{aff39}}
\and M.~Meneghetti\orcid{0000-0003-1225-7084}\inst{\ref{aff35},\ref{aff36}}
\and E.~Merlin\orcid{0000-0001-6870-8900}\inst{\ref{aff13}}
\and G.~Meylan\inst{\ref{aff102}}
\and A.~Mora\orcid{0000-0002-1922-8529}\inst{\ref{aff103}}
\and L.~Moscardini\orcid{0000-0002-3473-6716}\inst{\ref{aff42},\ref{aff35},\ref{aff36}}
\and E.~Munari\orcid{0000-0002-1751-5946}\inst{\ref{aff25},\ref{aff26}}
\and R.~Nakajima\orcid{0009-0009-1213-7040}\inst{\ref{aff97}}
\and C.~Neissner\orcid{0000-0001-8524-4968}\inst{\ref{aff104},\ref{aff54}}
\and S.-M.~Niemi\orcid{0009-0005-0247-0086}\inst{\ref{aff4}}
\and C.~Padilla\orcid{0000-0001-7951-0166}\inst{\ref{aff104}}
\and S.~Paltani\orcid{0000-0002-8108-9179}\inst{\ref{aff70}}
\and F.~Pasian\orcid{0000-0002-4869-3227}\inst{\ref{aff25}}
\and K.~Pedersen\inst{\ref{aff105}}
\and W.~J.~Percival\orcid{0000-0002-0644-5727}\inst{\ref{aff8},\ref{aff106},\ref{aff107}}
\and S.~Pires\orcid{0000-0002-0249-2104}\inst{\ref{aff19}}
\and G.~Polenta\orcid{0000-0003-4067-9196}\inst{\ref{aff76}}
\and M.~Poncet\inst{\ref{aff108}}
\and L.~A.~Popa\inst{\ref{aff109}}
\and L.~Pozzetti\orcid{0000-0001-7085-0412}\inst{\ref{aff35}}
\and F.~Raison\orcid{0000-0002-7819-6918}\inst{\ref{aff32}}
\and A.~Renzi\orcid{0000-0001-9856-1970}\inst{\ref{aff110},\ref{aff73}}
\and J.~Rhodes\orcid{0000-0002-4485-8549}\inst{\ref{aff82}}
\and G.~Riccio\inst{\ref{aff53}}
\and E.~Romelli\orcid{0000-0003-3069-9222}\inst{\ref{aff25}}
\and M.~Roncarelli\orcid{0000-0001-9587-7822}\inst{\ref{aff35}}
\and C.~Rosset\orcid{0000-0003-0286-2192}\inst{\ref{aff99}}
\and R.~Saglia\orcid{0000-0003-0378-7032}\inst{\ref{aff80},\ref{aff32}}
\and B.~Sartoris\orcid{0000-0003-1337-5269}\inst{\ref{aff80},\ref{aff25}}
\and P.~Schneider\orcid{0000-0001-8561-2679}\inst{\ref{aff97}}
\and T.~Schrabback\orcid{0000-0002-6987-7834}\inst{\ref{aff111}}
\and A.~Secroun\orcid{0000-0003-0505-3710}\inst{\ref{aff74}}
\and E.~Sefusatti\orcid{0000-0003-0473-1567}\inst{\ref{aff25},\ref{aff26},\ref{aff27}}
\and G.~Seidel\orcid{0000-0003-2907-353X}\inst{\ref{aff87}}
\and S.~Serrano\orcid{0000-0002-0211-2861}\inst{\ref{aff20},\ref{aff112},\ref{aff21}}
\and P.~Simon\inst{\ref{aff97}}
\and C.~Sirignano\orcid{0000-0002-0995-7146}\inst{\ref{aff110},\ref{aff73}}
\and G.~Sirri\orcid{0000-0003-2626-2853}\inst{\ref{aff36}}
\and A.~Spurio~Mancini\orcid{0000-0001-5698-0990}\inst{\ref{aff113}}
\and L.~Stanco\orcid{0000-0002-9706-5104}\inst{\ref{aff73}}
\and J.-L.~Starck\orcid{0000-0003-2177-7794}\inst{\ref{aff19}}
\and J.~Steinwagner\orcid{0000-0001-7443-1047}\inst{\ref{aff32}}
\and C.~Surace\orcid{0000-0003-2592-0113}\inst{\ref{aff71}}
\and P.~Tallada-Cresp\'{i}\orcid{0000-0002-1336-8328}\inst{\ref{aff24},\ref{aff54}}
\and A.~N.~Taylor\inst{\ref{aff7}}
\and I.~Tereno\orcid{0000-0002-4537-6218}\inst{\ref{aff68},\ref{aff114}}
\and N.~Tessore\orcid{0000-0002-9696-7931}\inst{\ref{aff67}}
\and S.~Toft\orcid{0000-0003-3631-7176}\inst{\ref{aff115},\ref{aff116}}
\and R.~Toledo-Moreo\orcid{0000-0002-2997-4859}\inst{\ref{aff117}}
\and F.~Torradeflot\orcid{0000-0003-1160-1517}\inst{\ref{aff54},\ref{aff24}}
\and E.~A.~Valentijn\inst{\ref{aff118}}
\and L.~Valenziano\orcid{0000-0002-1170-0104}\inst{\ref{aff35},\ref{aff77}}
\and J.~Valiviita\orcid{0000-0001-6225-3693}\inst{\ref{aff91},\ref{aff92}}
\and T.~Vassallo\orcid{0000-0001-6512-6358}\inst{\ref{aff25},\ref{aff78}}
\and A.~Veropalumbo\orcid{0000-0003-2387-1194}\inst{\ref{aff29},\ref{aff17},\ref{aff18}}
\and Y.~Wang\orcid{0000-0002-4749-2984}\inst{\ref{aff119}}
\and J.~Weller\orcid{0000-0002-8282-2010}\inst{\ref{aff80},\ref{aff32}}
\and A.~Zacchei\orcid{0000-0003-0396-1192}\inst{\ref{aff25},\ref{aff26}}
\and G.~Zamorani\orcid{0000-0002-2318-301X}\inst{\ref{aff35}}
\and F.~M.~Zerbi\inst{\ref{aff29}}
\and E.~Zucca\orcid{0000-0002-5845-8132}\inst{\ref{aff35}}
\and M.~Ballardini\orcid{0000-0003-4481-3559}\inst{\ref{aff40},\ref{aff41},\ref{aff35}}
\and M.~Bolzonella\orcid{0000-0003-3278-4607}\inst{\ref{aff35}}
\and A.~Boucaud\orcid{0000-0001-7387-2633}\inst{\ref{aff99}}
\and E.~Bozzo\orcid{0000-0002-8201-1525}\inst{\ref{aff70}}
\and C.~Burigana\orcid{0000-0002-3005-5796}\inst{\ref{aff120},\ref{aff77}}
\and R.~Cabanac\orcid{0000-0001-6679-2600}\inst{\ref{aff23}}
\and M.~Calabrese\orcid{0000-0002-2637-2422}\inst{\ref{aff121},\ref{aff9}}
\and A.~Cappi\inst{\ref{aff122},\ref{aff35}}
\and J.~A.~Escartin~Vigo\inst{\ref{aff32}}
\and L.~Gabarra\orcid{0000-0002-8486-8856}\inst{\ref{aff3}}
\and J.~Garc\'ia-Bellido\orcid{0000-0002-9370-8360}\inst{\ref{aff123}}
\and W.~G.~Hartley\inst{\ref{aff70}}
\and R.~Maoli\orcid{0000-0002-6065-3025}\inst{\ref{aff124},\ref{aff13}}
\and J.~Mart\'{i}n-Fleitas\orcid{0000-0002-8594-569X}\inst{\ref{aff125}}
\and M.~Maturi\orcid{0000-0002-3517-2422}\inst{\ref{aff30},\ref{aff126}}
\and N.~Mauri\orcid{0000-0001-8196-1548}\inst{\ref{aff57},\ref{aff36}}
\and R.~B.~Metcalf\orcid{0000-0003-3167-2574}\inst{\ref{aff42},\ref{aff35}}
\and M.~P\"ontinen\orcid{0000-0001-5442-2530}\inst{\ref{aff91}}
\and I.~Risso\orcid{0000-0003-2525-7761}\inst{\ref{aff29},\ref{aff17}}
\and V.~Scottez\orcid{0009-0008-3864-940X}\inst{\ref{aff101},\ref{aff127}}
\and M.~Sereno\orcid{0000-0003-0302-0325}\inst{\ref{aff35},\ref{aff36}}
\and M.~Tenti\orcid{0000-0002-4254-5901}\inst{\ref{aff36}}
\and M.~Viel\orcid{0000-0002-2642-5707}\inst{\ref{aff26},\ref{aff25},\ref{aff28},\ref{aff27},\ref{aff128}}
\and M.~Wiesmann\orcid{0009-0000-8199-5860}\inst{\ref{aff81}}
\and Y.~Akrami\orcid{0000-0002-2407-7956}\inst{\ref{aff123},\ref{aff129}}
\and S.~Alvi\orcid{0000-0001-5779-8568}\inst{\ref{aff40}}
\and I.~T.~Andika\orcid{0000-0001-6102-9526}\inst{\ref{aff130},\ref{aff131}}
\and S.~Anselmi\orcid{0000-0002-3579-9583}\inst{\ref{aff73},\ref{aff110},\ref{aff132}}
\and M.~Archidiacono\orcid{0000-0003-4952-9012}\inst{\ref{aff95},\ref{aff96}}
\and F.~Atrio-Barandela\orcid{0000-0002-2130-2513}\inst{\ref{aff133}}
\and E.~Aubourg\orcid{0000-0002-5592-023X}\inst{\ref{aff99},\ref{aff134}}
\and L.~Bazzanini\orcid{0000-0003-0727-0137}\inst{\ref{aff40},\ref{aff35}}
\and M.~Bethermin\orcid{0000-0002-3915-2015}\inst{\ref{aff135}}
\and A.~Blanchard\orcid{0000-0001-8555-9003}\inst{\ref{aff23}}
\and S.~Borgani\orcid{0000-0001-6151-6439}\inst{\ref{aff136},\ref{aff26},\ref{aff25},\ref{aff27},\ref{aff128}}
\and M.~L.~Brown\orcid{0000-0002-0370-8077}\inst{\ref{aff59}}
\and S.~Bruton\orcid{0000-0002-6503-5218}\inst{\ref{aff137}}
\and A.~Calabro\orcid{0000-0003-2536-1614}\inst{\ref{aff13}}
\and F.~Caro\inst{\ref{aff13}}
\and C.~S.~Carvalho\inst{\ref{aff114}}
\and T.~Castro\orcid{0000-0002-6292-3228}\inst{\ref{aff25},\ref{aff27},\ref{aff26},\ref{aff128}}
\and F.~Cogato\orcid{0000-0003-4632-6113}\inst{\ref{aff42},\ref{aff35}}
\and S.~Conseil\orcid{0000-0002-3657-4191}\inst{\ref{aff62}}
\and O.~Cucciati\orcid{0000-0002-9336-7551}\inst{\ref{aff35}}
\and G.~Desprez\orcid{0000-0001-8325-1742}\inst{\ref{aff118}}
\and A.~D\'iaz-S\'anchez\orcid{0000-0003-0748-4768}\inst{\ref{aff138}}
\and J.~M.~Diego\orcid{0000-0001-9065-3926}\inst{\ref{aff44}}
\and M.~Y.~Elkhashab\orcid{0000-0001-9306-2603}\inst{\ref{aff25},\ref{aff27},\ref{aff136},\ref{aff26}}
\and Y.~Fang\orcid{0000-0002-0334-6950}\inst{\ref{aff80}}
\and A.~G.~Ferrari\orcid{0009-0005-5266-4110}\inst{\ref{aff36}}
\and P.~G.~Ferreira\orcid{0000-0002-3021-2851}\inst{\ref{aff3}}
\and A.~Finoguenov\orcid{0000-0002-4606-5403}\inst{\ref{aff91}}
\and A.~Franco\orcid{0000-0002-4761-366X}\inst{\ref{aff139},\ref{aff140},\ref{aff141}}
\and K.~Ganga\orcid{0000-0001-8159-8208}\inst{\ref{aff99}}
\and T.~Gasparetto\orcid{0000-0002-7913-4866}\inst{\ref{aff13}}
\and V.~Gautard\inst{\ref{aff142}}
\and R.~Gavazzi\orcid{0000-0002-5540-6935}\inst{\ref{aff71},\ref{aff39}}
\and E.~Gaztanaga\orcid{0000-0001-9632-0815}\inst{\ref{aff21},\ref{aff20},\ref{aff143}}
\and F.~Giacomini\orcid{0000-0002-3129-2814}\inst{\ref{aff36}}
\and F.~Gianotti\orcid{0000-0003-4666-119X}\inst{\ref{aff35}}
\and G.~Gozaliasl\orcid{0000-0002-0236-919X}\inst{\ref{aff144},\ref{aff91}}
\and A.~Gruppuso\orcid{0000-0001-9272-5292}\inst{\ref{aff35},\ref{aff36}}
\and M.~Guidi\orcid{0000-0001-9408-1101}\inst{\ref{aff51},\ref{aff35}}
\and C.~M.~Gutierrez\orcid{0000-0001-7854-783X}\inst{\ref{aff145}}
\and H.~Hildebrandt\orcid{0000-0002-9814-3338}\inst{\ref{aff146}}
\and J.~Hjorth\orcid{0000-0002-4571-2306}\inst{\ref{aff105}}
\and J.~J.~E.~Kajava\orcid{0000-0002-3010-8333}\inst{\ref{aff147},\ref{aff148}}
\and Y.~Kang\orcid{0009-0000-8588-7250}\inst{\ref{aff70}}
\and V.~Kansal\orcid{0000-0002-4008-6078}\inst{\ref{aff149},\ref{aff150}}
\and D.~Karagiannis\orcid{0000-0002-4927-0816}\inst{\ref{aff40},\ref{aff151}}
\and K.~Kiiveri\inst{\ref{aff90}}
\and J.~Kim\orcid{0000-0003-2776-2761}\inst{\ref{aff3}}
\and C.~C.~Kirkpatrick\inst{\ref{aff90}}
\and S.~Kruk\orcid{0000-0001-8010-8879}\inst{\ref{aff61}}
\and M.~Lattanzi\orcid{0000-0003-1059-2532}\inst{\ref{aff41}}
\and V.~Le~Brun\orcid{0000-0002-5027-1939}\inst{\ref{aff71}}
\and F.~Lepori\orcid{0009-0000-5061-7138}\inst{\ref{aff152}}
\and G.~Leroy\orcid{0009-0004-2523-4425}\inst{\ref{aff153},\ref{aff98}}
\and J.~Lesgourgues\orcid{0000-0001-7627-353X}\inst{\ref{aff15}}
\and L.~Leuzzi\orcid{0009-0006-4479-7017}\inst{\ref{aff35}}
\and T.~I.~Liaudat\orcid{0000-0002-9104-314X}\inst{\ref{aff134}}
\and J.~Macias-Perez\orcid{0000-0002-5385-2763}\inst{\ref{aff154}}
\and M.~Magliocchetti\orcid{0000-0001-9158-4838}\inst{\ref{aff75}}
\and F.~Mannucci\orcid{0000-0002-4803-2381}\inst{\ref{aff155}}
\and C.~J.~A.~P.~Martins\orcid{0000-0002-4886-9261}\inst{\ref{aff156},\ref{aff157}}
\and L.~Maurin\orcid{0000-0002-8406-0857}\inst{\ref{aff72}}
\and M.~Migliaccio\inst{\ref{aff158},\ref{aff159}}
\and M.~Miluzio\inst{\ref{aff61},\ref{aff160}}
\and P.~Monaco\orcid{0000-0003-2083-7564}\inst{\ref{aff136},\ref{aff25},\ref{aff27},\ref{aff26}}
\and A.~Montoro\orcid{0000-0003-4730-8590}\inst{\ref{aff21},\ref{aff20}}
\and G.~Morgante\inst{\ref{aff35}}
\and S.~Nadathur\orcid{0000-0001-9070-3102}\inst{\ref{aff143}}
\and K.~Naidoo\orcid{0000-0002-9182-1802}\inst{\ref{aff143},\ref{aff89}}
\and A.~Navarro-Alsina\orcid{0000-0002-3173-2592}\inst{\ref{aff97}}
\and S.~Nesseris\orcid{0000-0002-0567-0324}\inst{\ref{aff123}}
\and L.~Pagano\orcid{0000-0003-1820-5998}\inst{\ref{aff40},\ref{aff41}}
\and D.~Paoletti\orcid{0000-0003-4761-6147}\inst{\ref{aff35},\ref{aff77}}
\and F.~Passalacqua\orcid{0000-0002-8606-4093}\inst{\ref{aff110},\ref{aff73}}
\and K.~Paterson\orcid{0000-0001-8340-3486}\inst{\ref{aff87}}
\and R.~Paviot\orcid{0009-0002-8108-3460}\inst{\ref{aff19}}
\and A.~Pisani\orcid{0000-0002-6146-4437}\inst{\ref{aff74}}
\and D.~Potter\orcid{0000-0002-0757-5195}\inst{\ref{aff152}}
\and S.~Quai\orcid{0000-0002-0449-8163}\inst{\ref{aff42},\ref{aff35}}
\and M.~Radovich\orcid{0000-0002-3585-866X}\inst{\ref{aff79}}
\and W.~Roster\orcid{0000-0002-9149-6528}\inst{\ref{aff32}}
\and S.~Sacquegna\orcid{0000-0002-8433-6630}\inst{\ref{aff161}}
\and M.~Sahl\'en\orcid{0000-0003-0973-4804}\inst{\ref{aff162}}
\and D.~B.~Sanders\orcid{0000-0002-1233-9998}\inst{\ref{aff56}}
\and E.~Sarpa\orcid{0000-0002-1256-655X}\inst{\ref{aff28},\ref{aff128},\ref{aff27}}
\and J.~Schaye\orcid{0000-0002-0668-5560}\inst{\ref{aff5}}
\and A.~Schneider\orcid{0000-0001-7055-8104}\inst{\ref{aff152}}
\and L.~C.~Smith\orcid{0000-0002-3259-2771}\inst{\ref{aff163}}
\and J.~G.~Sorce\orcid{0000-0002-2307-2432}\inst{\ref{aff164},\ref{aff72}}
\and J.~Stadel\orcid{0000-0001-7565-8622}\inst{\ref{aff152}}
\and C.~Tao\orcid{0000-0001-7961-8177}\inst{\ref{aff74}}
\and G.~Testera\inst{\ref{aff17}}
\and R.~Teyssier\orcid{0000-0001-7689-0933}\inst{\ref{aff165}}
\and S.~Tosi\orcid{0000-0002-7275-9193}\inst{\ref{aff18},\ref{aff17},\ref{aff29}}
\and A.~Troja\orcid{0000-0003-0239-4595}\inst{\ref{aff110},\ref{aff73}}
\and M.~Tucci\inst{\ref{aff70}}
\and A.~Venhola\orcid{0000-0001-6071-4564}\inst{\ref{aff166}}
\and D.~Vergani\orcid{0000-0003-0898-2216}\inst{\ref{aff35}}
\and F.~Vernizzi\orcid{0000-0003-3426-2802}\inst{\ref{aff167}}
\and G.~Verza\orcid{0000-0002-1886-8348}\inst{\ref{aff168},\ref{aff169}}
\and S.~Vinciguerra\orcid{0009-0005-4018-3184}\inst{\ref{aff71}}
\and N.~A.~Walton\orcid{0000-0003-3983-8778}\inst{\ref{aff163}}}
										   
\institute{Center for Data-Driven Discovery, Kavli IPMU (WPI), UTIAS, The University of Tokyo, Kashiwa, Chiba 277-8583, Japan\label{aff1}
\and
Laboratoire d'etude de l'Univers et des phenomenes eXtremes, Observatoire de Paris, Universit\'e PSL, Sorbonne Universit\'e, CNRS, 92190 Meudon, France\label{aff2}
\and
Department of Physics, Oxford University, Keble Road, Oxford OX1 3RH, UK\label{aff3}
\and
European Space Agency/ESTEC, Keplerlaan 1, 2201 AZ Noordwijk, The Netherlands\label{aff4}
\and
Leiden Observatory, Leiden University, Einsteinweg 55, 2333 CC Leiden, The Netherlands\label{aff5}
\and
Department of Physics, Astronomy and Mathematics, University of Hertfordshire, College Lane, Hatfield AL10 9AB, UK\label{aff6}
\and
Institute for Astronomy, University of Edinburgh, Royal Observatory, Blackford Hill, Edinburgh EH9 3HJ, UK\label{aff7}
\and
Waterloo Centre for Astrophysics, University of Waterloo, Waterloo, Ontario N2L 3G1, Canada\label{aff8}
\and
INAF-IASF Milano, Via Alfonso Corti 12, 20133 Milano, Italy\label{aff9}
\and
Dipartimento di Fisica, Universit\`a degli Studi di Torino, Via P. Giuria 1, 10125 Torino, Italy\label{aff10}
\and
INFN-Sezione di Torino, Via P. Giuria 1, 10125 Torino, Italy\label{aff11}
\and
INAF-Osservatorio Astrofisico di Torino, Via Osservatorio 20, 10025 Pino Torinese (TO), Italy\label{aff12}
\and
INAF-Osservatorio Astronomico di Roma, Via Frascati 33, 00078 Monteporzio Catone, Italy\label{aff13}
\and
INFN-Sezione di Roma, Piazzale Aldo Moro, 2 - c/o Dipartimento di Fisica, Edificio G. Marconi, 00185 Roma, Italy\label{aff14}
\and
Institute for Theoretical Particle Physics and Cosmology (TTK), RWTH Aachen University, 52056 Aachen, Germany\label{aff15}
\and
Deutsches Zentrum f\"ur Luft- und Raumfahrt e. V. (DLR), Linder H\"ohe, 51147 K\"oln, Germany\label{aff16}
\and
INFN-Sezione di Genova, Via Dodecaneso 33, 16146, Genova, Italy\label{aff17}
\and
Dipartimento di Fisica, Universit\`a di Genova, Via Dodecaneso 33, 16146, Genova, Italy\label{aff18}
\and
Universit\'e Paris-Saclay, Universit\'e Paris Cit\'e, CEA, CNRS, AIM, 91191, Gif-sur-Yvette, France\label{aff19}
\and
Institut d'Estudis Espacials de Catalunya (IEEC),  Edifici RDIT, Campus UPC, 08860 Castelldefels, Barcelona, Spain\label{aff20}
\and
Institute of Space Sciences (ICE, CSIC), Campus UAB, Carrer de Can Magrans, s/n, 08193 Barcelona, Spain\label{aff21}
\and
Universit\'e Paris-Saclay, CNRS/IN2P3, IJCLab, 91405 Orsay, France\label{aff22}
\and
Institut de Recherche en Astrophysique et Plan\'etologie (IRAP), Universit\'e de Toulouse, CNRS, UPS, CNES, 14 Av. Edouard Belin, 31400 Toulouse, France\label{aff23}
\and
Centro de Investigaciones Energ\'eticas, Medioambientales y Tecnol\'ogicas (CIEMAT), Avenida Complutense 40, 28040 Madrid, Spain\label{aff24}
\and
INAF-Osservatorio Astronomico di Trieste, Via G. B. Tiepolo 11, 34143 Trieste, Italy\label{aff25}
\and
IFPU, Institute for Fundamental Physics of the Universe, via Beirut 2, 34151 Trieste, Italy\label{aff26}
\and
INFN, Sezione di Trieste, Via Valerio 2, 34127 Trieste TS, Italy\label{aff27}
\and
SISSA, International School for Advanced Studies, Via Bonomea 265, 34136 Trieste TS, Italy\label{aff28}
\and
INAF-Osservatorio Astronomico di Brera, Via Brera 28, 20122 Milano, Italy\label{aff29}
\and
Institut f\"ur Theoretische Physik, University of Heidelberg, Philosophenweg 16, 69120 Heidelberg, Germany\label{aff30}
\and
Universit\'e St Joseph; Faculty of Sciences, Beirut, Lebanon\label{aff31}
\and
Max Planck Institute for Extraterrestrial Physics, Giessenbachstr. 1, 85748 Garching, Germany\label{aff32}
\and
LINKS Foundation, Via Pier Carlo Boggio, 61 10138 Torino, Italy\label{aff33}
\and
Institute for Particle Physics and Astrophysics, Dept. of Physics, ETH Zurich, Wolfgang-Pauli-Strasse 27, 8093 Zurich, Switzerland\label{aff34}
\and
INAF-Osservatorio di Astrofisica e Scienza dello Spazio di Bologna, Via Piero Gobetti 93/3, 40129 Bologna, Italy\label{aff35}
\and
INFN-Sezione di Bologna, Viale Berti Pichat 6/2, 40127 Bologna, Italy\label{aff36}
\and
DAMTP, Centre for Mathematical Sciences, Wilberforce Road, Cambridge CB3 0WA, UK\label{aff37}
\and
Kavli Institute for Cosmology Cambridge, Madingley Road, Cambridge, CB3 0HA, UK\label{aff38}
\and
Institut d'Astrophysique de Paris, UMR 7095, CNRS, and Sorbonne Universit\'e, 98 bis boulevard Arago, 75014 Paris, France\label{aff39}
\and
Dipartimento di Fisica e Scienze della Terra, Universit\`a degli Studi di Ferrara, Via Giuseppe Saragat 1, 44122 Ferrara, Italy\label{aff40}
\and
Istituto Nazionale di Fisica Nucleare, Sezione di Ferrara, Via Giuseppe Saragat 1, 44122 Ferrara, Italy\label{aff41}
\and
Dipartimento di Fisica e Astronomia "Augusto Righi" - Alma Mater Studiorum Universit\`a di Bologna, via Piero Gobetti 93/2, 40129 Bologna, Italy\label{aff42}
\and
Universit\'e de Gen\`eve, D\'epartement de Physique Th\'eorique and Centre for Astroparticle Physics, 24 quai Ernest-Ansermet, CH-1211 Gen\`eve 4, Switzerland\label{aff43}
\and
Instituto de F\'isica de Cantabria, Edificio Juan Jord\'a, Avenida de los Castros, 39005 Santander, Spain\label{aff44}
\and
Higgs Centre for Theoretical Physics, School of Physics and Astronomy, The University of Edinburgh, Edinburgh EH9 3FD, UK\label{aff45}
\and
Aix-Marseille Universit\'e, Universit\'e de Toulon, CNRS, CPT, Marseille, France\label{aff46}
\and
Departamento de F\'isica, FCFM, Universidad de Chile, Blanco Encalada 2008, Santiago, Chile\label{aff47}
\and
Mathematical Institute, University of Leiden, Einsteinweg 55, 2333 CA Leiden, The Netherlands\label{aff48}
\and
Center for Cosmology and AstroParticle Physics, The Ohio State University, 191 West Woodruff Avenue, Columbus, OH 43210, USA\label{aff49}
\and
Department of Physics, The Ohio State University, Columbus, OH 43210, USA\label{aff50}
\and
Dipartimento di Fisica e Astronomia, Universit\`a di Bologna, Via Gobetti 93/2, 40129 Bologna, Italy\label{aff51}
\and
Department of Physics "E. Pancini", University Federico II, Via Cinthia 6, 80126, Napoli, Italy\label{aff52}
\and
INAF-Osservatorio Astronomico di Capodimonte, Via Moiariello 16, 80131 Napoli, Italy\label{aff53}
\and
Port d'Informaci\'{o} Cient\'{i}fica, Campus UAB, C. Albareda s/n, 08193 Bellaterra (Barcelona), Spain\label{aff54}
\and
INFN section of Naples, Via Cinthia 6, 80126, Napoli, Italy\label{aff55}
\and
Institute for Astronomy, University of Hawaii, 2680 Woodlawn Drive, Honolulu, HI 96822, USA\label{aff56}
\and
Dipartimento di Fisica e Astronomia "Augusto Righi" - Alma Mater Studiorum Universit\`a di Bologna, Viale Berti Pichat 6/2, 40127 Bologna, Italy\label{aff57}
\and
Instituto de Astrof\'{\i}sica de Canarias, V\'{\i}a L\'actea, 38205 La Laguna, Tenerife, Spain\label{aff58}
\and
Jodrell Bank Centre for Astrophysics, Department of Physics and Astronomy, University of Manchester, Oxford Road, Manchester M13 9PL, UK\label{aff59}
\and
European Space Agency/ESRIN, Largo Galileo Galilei 1, 00044 Frascati, Roma, Italy\label{aff60}
\and
ESAC/ESA, Camino Bajo del Castillo, s/n., Urb. Villafranca del Castillo, 28692 Villanueva de la Ca\~nada, Madrid, Spain\label{aff61}
\and
Universit\'e Claude Bernard Lyon 1, CNRS/IN2P3, IP2I Lyon, UMR 5822, Villeurbanne, F-69100, France\label{aff62}
\and
Institut de Ci\`{e}ncies del Cosmos (ICCUB), Universitat de Barcelona (IEEC-UB), Mart\'{i} i Franqu\`{e}s 1, 08028 Barcelona, Spain\label{aff63}
\and
Instituci\'o Catalana de Recerca i Estudis Avan\c{c}ats (ICREA), Passeig de Llu\'{\i}s Companys 23, 08010 Barcelona, Spain\label{aff64}
\and
Institut de Ciencies de l'Espai (IEEC-CSIC), Campus UAB, Carrer de Can Magrans, s/n Cerdanyola del Vall\'es, 08193 Barcelona, Spain\label{aff65}
\and
UCB Lyon 1, CNRS/IN2P3, IUF, IP2I Lyon, 4 rue Enrico Fermi, 69622 Villeurbanne, France\label{aff66}
\and
Mullard Space Science Laboratory, University College London, Holmbury St Mary, Dorking, Surrey RH5 6NT, UK\label{aff67}
\and
Departamento de F\'isica, Faculdade de Ci\^encias, Universidade de Lisboa, Edif\'icio C8, Campo Grande, PT1749-016 Lisboa, Portugal\label{aff68}
\and
Instituto de Astrof\'isica e Ci\^encias do Espa\c{c}o, Faculdade de Ci\^encias, Universidade de Lisboa, Campo Grande, 1749-016 Lisboa, Portugal\label{aff69}
\and
Department of Astronomy, University of Geneva, ch. d'Ecogia 16, 1290 Versoix, Switzerland\label{aff70}
\and
Aix-Marseille Universit\'e, CNRS, CNES, LAM, Marseille, France\label{aff71}
\and
Universit\'e Paris-Saclay, CNRS, Institut d'astrophysique spatiale, 91405, Orsay, France\label{aff72}
\and
INFN-Padova, Via Marzolo 8, 35131 Padova, Italy\label{aff73}
\and
Aix-Marseille Universit\'e, CNRS/IN2P3, CPPM, Marseille, France\label{aff74}
\and
INAF-Istituto di Astrofisica e Planetologia Spaziali, via del Fosso del Cavaliere, 100, 00100 Roma, Italy\label{aff75}
\and
Space Science Data Center, Italian Space Agency, via del Politecnico snc, 00133 Roma, Italy\label{aff76}
\and
INFN-Bologna, Via Irnerio 46, 40126 Bologna, Italy\label{aff77}
\and
University Observatory, LMU Faculty of Physics, Scheinerstrasse 1, 81679 Munich, Germany\label{aff78}
\and
INAF-Osservatorio Astronomico di Padova, Via dell'Osservatorio 5, 35122 Padova, Italy\label{aff79}
\and
Universit\"ats-Sternwarte M\"unchen, Fakult\"at f\"ur Physik, Ludwig-Maximilians-Universit\"at M\"unchen, Scheinerstrasse 1, 81679 M\"unchen, Germany\label{aff80}
\and
Institute of Theoretical Astrophysics, University of Oslo, P.O. Box 1029 Blindern, 0315 Oslo, Norway\label{aff81}
\and
Jet Propulsion Laboratory, California Institute of Technology, 4800 Oak Grove Drive, Pasadena, CA, 91109, USA\label{aff82}
\and
Department of Physics, Lancaster University, Lancaster, LA1 4YB, UK\label{aff83}
\and
Felix Hormuth Engineering, Goethestr. 17, 69181 Leimen, Germany\label{aff84}
\and
Technical University of Denmark, Elektrovej 327, 2800 Kgs. Lyngby, Denmark\label{aff85}
\and
Cosmic Dawn Center (DAWN), Denmark\label{aff86}
\and
Max-Planck-Institut f\"ur Astronomie, K\"onigstuhl 17, 69117 Heidelberg, Germany\label{aff87}
\and
NASA Goddard Space Flight Center, Greenbelt, MD 20771, USA\label{aff88}
\and
Department of Physics and Astronomy, University College London, Gower Street, London WC1E 6BT, UK\label{aff89}
\and
Department of Physics and Helsinki Institute of Physics, Gustaf H\"allstr\"omin katu 2, University of Helsinki, 00014 Helsinki, Finland\label{aff90}
\and
Department of Physics, P.O. Box 64, University of Helsinki, 00014 Helsinki, Finland\label{aff91}
\and
Helsinki Institute of Physics, Gustaf H{\"a}llstr{\"o}min katu 2, University of Helsinki, 00014 Helsinki, Finland\label{aff92}
\and
SKAO, Jodrell Bank, Lower Withington, Macclesfield SK11 9FT, UK\label{aff93}
\and
Centre de Calcul de l'IN2P3/CNRS, 21 avenue Pierre de Coubertin 69627 Villeurbanne Cedex, France\label{aff94}
\and
Dipartimento di Fisica "Aldo Pontremoli", Universit\`a degli Studi di Milano, Via Celoria 16, 20133 Milano, Italy\label{aff95}
\and
INFN-Sezione di Milano, Via Celoria 16, 20133 Milano, Italy\label{aff96}
\and
Universit\"at Bonn, Argelander-Institut f\"ur Astronomie, Auf dem H\"ugel 71, 53121 Bonn, Germany\label{aff97}
\and
Department of Physics, Institute for Computational Cosmology, Durham University, South Road, Durham, DH1 3LE, UK\label{aff98}
\and
Universit\'e Paris Cit\'e, CNRS, Astroparticule et Cosmologie, 75013 Paris, France\label{aff99}
\and
CNRS-UCB International Research Laboratory, Centre Pierre Bin\'etruy, IRL2007, CPB-IN2P3, Berkeley, USA\label{aff100}
\and
Institut d'Astrophysique de Paris, 98bis Boulevard Arago, 75014, Paris, France\label{aff101}
\and
Institute of Physics, Laboratory of Astrophysics, Ecole Polytechnique F\'ed\'erale de Lausanne (EPFL), Observatoire de Sauverny, 1290 Versoix, Switzerland\label{aff102}
\and
Telespazio UK S.L. for European Space Agency (ESA), Camino bajo del Castillo, s/n, Urbanizacion Villafranca del Castillo, Villanueva de la Ca\~nada, 28692 Madrid, Spain\label{aff103}
\and
Institut de F\'{i}sica d'Altes Energies (IFAE), The Barcelona Institute of Science and Technology, Campus UAB, 08193 Bellaterra (Barcelona), Spain\label{aff104}
\and
DARK, Niels Bohr Institute, University of Copenhagen, Jagtvej 155, 2200 Copenhagen, Denmark\label{aff105}
\and
Department of Physics and Astronomy, University of Waterloo, Waterloo, Ontario N2L 3G1, Canada\label{aff106}
\and
Perimeter Institute for Theoretical Physics, Waterloo, Ontario N2L 2Y5, Canada\label{aff107}
\and
Centre National d'Etudes Spatiales -- Centre spatial de Toulouse, 18 avenue Edouard Belin, 31401 Toulouse Cedex 9, France\label{aff108}
\and
Institute of Space Science, Str. Atomistilor, nr. 409 M\u{a}gurele, Ilfov, 077125, Romania\label{aff109}
\and
Dipartimento di Fisica e Astronomia "G. Galilei", Universit\`a di Padova, Via Marzolo 8, 35131 Padova, Italy\label{aff110}
\and
Universit\"at Innsbruck, Institut f\"ur Astro- und Teilchenphysik, Technikerstr. 25/8, 6020 Innsbruck, Austria\label{aff111}
\and
Satlantis, University Science Park, Sede Bld 48940, Leioa-Bilbao, Spain\label{aff112}
\and
Department of Physics, Royal Holloway, University of London, Surrey TW20 0EX, UK\label{aff113}
\and
Instituto de Astrof\'isica e Ci\^encias do Espa\c{c}o, Faculdade de Ci\^encias, Universidade de Lisboa, Tapada da Ajuda, 1349-018 Lisboa, Portugal\label{aff114}
\and
Cosmic Dawn Center (DAWN)\label{aff115}
\and
Niels Bohr Institute, University of Copenhagen, Jagtvej 128, 2200 Copenhagen, Denmark\label{aff116}
\and
Universidad Polit\'ecnica de Cartagena, Departamento de Electr\'onica y Tecnolog\'ia de Computadoras,  Plaza del Hospital 1, 30202 Cartagena, Spain\label{aff117}
\and
Kapteyn Astronomical Institute, University of Groningen, PO Box 800, 9700 AV Groningen, The Netherlands\label{aff118}
\and
Infrared Processing and Analysis Center, California Institute of Technology, Pasadena, CA 91125, USA\label{aff119}
\and
INAF, Istituto di Radioastronomia, Via Piero Gobetti 101, 40129 Bologna, Italy\label{aff120}
\and
Astronomical Observatory of the Autonomous Region of the Aosta Valley (OAVdA), Loc. Lignan 39, I-11020, Nus (Aosta Valley), Italy\label{aff121}
\and
Universit\'e C\^{o}te d'Azur, Observatoire de la C\^{o}te d'Azur, CNRS, Laboratoire Lagrange, Bd de l'Observatoire, CS 34229, 06304 Nice cedex 4, France\label{aff122}
\and
Instituto de F\'isica Te\'orica UAM-CSIC, Campus de Cantoblanco, 28049 Madrid, Spain\label{aff123}
\and
Dipartimento di Fisica, Sapienza Universit\`a di Roma, Piazzale Aldo Moro 2, 00185 Roma, Italy\label{aff124}
\and
Aurora Technology for European Space Agency (ESA), Camino bajo del Castillo, s/n, Urbanizacion Villafranca del Castillo, Villanueva de la Ca\~nada, 28692 Madrid, Spain\label{aff125}
\and
Zentrum f\"ur Astronomie, Universit\"at Heidelberg, Philosophenweg 12, 69120 Heidelberg, Germany\label{aff126}
\and
ICL, Junia, Universit\'e Catholique de Lille, LITL, 59000 Lille, France\label{aff127}
\and
ICSC - Centro Nazionale di Ricerca in High Performance Computing, Big Data e Quantum Computing, Via Magnanelli 2, Bologna, Italy\label{aff128}
\and
CERCA/ISO, Department of Physics, Case Western Reserve University, 10900 Euclid Avenue, Cleveland, OH 44106, USA\label{aff129}
\and
Technical University of Munich, TUM School of Natural Sciences, Physics Department, James-Franck-Str.~1, 85748 Garching, Germany\label{aff130}
\and
Max-Planck-Institut f\"ur Astrophysik, Karl-Schwarzschild-Str.~1, 85748 Garching, Germany\label{aff131}
\and
Laboratoire Univers et Th\'eorie, Observatoire de Paris, Universit\'e PSL, Universit\'e Paris Cit\'e, CNRS, 92190 Meudon, France\label{aff132}
\and
Departamento de F{\'\i}sica Fundamental. Universidad de Salamanca. Plaza de la Merced s/n. 37008 Salamanca, Spain\label{aff133}
\and
IRFU, CEA, Universit\'e Paris-Saclay 91191 Gif-sur-Yvette Cedex, France\label{aff134}
\and
Universit\'e de Strasbourg, CNRS, Observatoire astronomique de Strasbourg, UMR 7550, 67000 Strasbourg, France\label{aff135}
\and
Dipartimento di Fisica - Sezione di Astronomia, Universit\`a di Trieste, Via Tiepolo 11, 34131 Trieste, Italy\label{aff136}
\and
California Institute of Technology, 1200 E California Blvd, Pasadena, CA 91125, USA\label{aff137}
\and
Departamento F\'isica Aplicada, Universidad Polit\'ecnica de Cartagena, Campus Muralla del Mar, 30202 Cartagena, Murcia, Spain\label{aff138}
\and
INFN, Sezione di Lecce, Via per Arnesano, CP-193, 73100, Lecce, Italy\label{aff139}
\and
Department of Mathematics and Physics E. De Giorgi, University of Salento, Via per Arnesano, CP-I93, 73100, Lecce, Italy\label{aff140}
\and
INAF-Sezione di Lecce, c/o Dipartimento Matematica e Fisica, Via per Arnesano, 73100, Lecce, Italy\label{aff141}
\and
CEA Saclay, DFR/IRFU, Service d'Astrophysique, Bat. 709, 91191 Gif-sur-Yvette, France\label{aff142}
\and
Institute of Cosmology and Gravitation, University of Portsmouth, Portsmouth PO1 3FX, UK\label{aff143}
\and
Department of Computer Science, Aalto University, PO Box 15400, Espoo, FI-00 076, Finland\label{aff144}
\and
Instituto de Astrof\'\i sica de Canarias, c/ Via Lactea s/n, La Laguna 38200, Spain. Departamento de Astrof\'\i sica de la Universidad de La Laguna, Avda. Francisco Sanchez, La Laguna, 38200, Spain\label{aff145}
\and
Ruhr University Bochum, Faculty of Physics and Astronomy, Astronomical Institute (AIRUB), German Centre for Cosmological Lensing (GCCL), 44780 Bochum, Germany\label{aff146}
\and
Department of Physics and Astronomy, Vesilinnantie 5, University of Turku, 20014 Turku, Finland\label{aff147}
\and
Serco for European Space Agency (ESA), Camino bajo del Castillo, s/n, Urbanizacion Villafranca del Castillo, Villanueva de la Ca\~nada, 28692 Madrid, Spain\label{aff148}
\and
ARC Centre of Excellence for Dark Matter Particle Physics, Melbourne, Australia\label{aff149}
\and
Centre for Astrophysics \& Supercomputing, Swinburne University of Technology,  Hawthorn, Victoria 3122, Australia\label{aff150}
\and
Department of Physics and Astronomy, University of the Western Cape, Bellville, Cape Town, 7535, South Africa\label{aff151}
\and
Department of Astrophysics, University of Zurich, Winterthurerstrasse 190, 8057 Zurich, Switzerland\label{aff152}
\and
Department of Physics, Centre for Extragalactic Astronomy, Durham University, South Road, Durham, DH1 3LE, UK\label{aff153}
\and
Univ. Grenoble Alpes, CNRS, Grenoble INP, LPSC-IN2P3, 53, Avenue des Martyrs, 38000, Grenoble, France\label{aff154}
\and
INAF-Osservatorio Astrofisico di Arcetri, Largo E. Fermi 5, 50125, Firenze, Italy\label{aff155}
\and
Centro de Astrof\'{\i}sica da Universidade do Porto, Rua das Estrelas, 4150-762 Porto, Portugal\label{aff156}
\and
Instituto de Astrof\'isica e Ci\^encias do Espa\c{c}o, Universidade do Porto, CAUP, Rua das Estrelas, PT4150-762 Porto, Portugal\label{aff157}
\and
Dipartimento di Fisica, Universit\`a di Roma Tor Vergata, Via della Ricerca Scientifica 1, Roma, Italy\label{aff158}
\and
INFN, Sezione di Roma 2, Via della Ricerca Scientifica 1, Roma, Italy\label{aff159}
\and
HE Space for European Space Agency (ESA), Camino bajo del Castillo, s/n, Urbanizacion Villafranca del Castillo, Villanueva de la Ca\~nada, 28692 Madrid, Spain\label{aff160}
\and
INAF - Osservatorio Astronomico d'Abruzzo, Via Maggini, 64100, Teramo, Italy\label{aff161}
\and
Theoretical astrophysics, Department of Physics and Astronomy, Uppsala University, Box 516, 751 37 Uppsala, Sweden\label{aff162}
\and
Institute of Astronomy, University of Cambridge, Madingley Road, Cambridge CB3 0HA, UK\label{aff163}
\and
Univ. Lille, CNRS, Centrale Lille, UMR 9189 CRIStAL, 59000 Lille, France\label{aff164}
\and
Department of Astrophysical Sciences, Peyton Hall, Princeton University, Princeton, NJ 08544, USA\label{aff165}
\and
Space physics and astronomy research unit, University of Oulu, Pentti Kaiteran katu 1, FI-90014 Oulu, Finland\label{aff166}
\and
Institut de Physique Th\'eorique, CEA, CNRS, Universit\'e Paris-Saclay 91191 Gif-sur-Yvette Cedex, France\label{aff167}
\and
Center for Computational Astrophysics, Flatiron Institute, 162 5th Avenue, 10010, New York, NY, USA\label{aff168}
\and
International Centre for Theoretical Physics (ICTP), Strada Costiera 11, 34151 Trieste, Italy\label{aff169}}    

\date{\today}

\authorrunning{Euclid Collaboration: L. Blot et al.}

\titlerunning{CLOE. 6. Impact of systematics}

  \abstract{Extracting cosmological information from the \Euclid galaxy survey will require modelling numerous systematic effects during the inference process. This implies varying a large number of nuisance parameters, which have to be marginalised over before reporting the constraints on the cosmological parameters. This is a delicate process, especially with such a large parameter space, which could result in biased cosmological results. In this work, we study the impact of different choices for modelling systematic effects and prior distribution of nuisance parameters for the final \Euclid Data Release, focusing on the 3$\times$2pt analysis for photometric probes and the galaxy power spectrum multipoles for the spectroscopic probes. We explore the effect of intrinsic alignments, linear galaxy bias, magnification bias, multiplicative cosmic shear bias and shifts in the redshift distribution for the photometric probes, as well as the purity of the spectroscopic sample. We find that intrinsic alignment modelling has the most severe impact with a bias up to $6\,\sigma$ on the Hubble constant $H_0$ if neglected, followed by mis-modelling of the redshift evolution of galaxy bias, yielding up to $1.5\,\sigma$ on the parameter $S_8\equiv\sotto\sqrt{\Om /0.3}$. Choosing a too optimistic prior for multiplicative bias can also result in biases of the order of $0.7\,\sigma$ on $S_8$. We also find that the precision on the estimate of the purity of the spectroscopic sample will be an important driver for the constraining power of the galaxy clustering full-shape analysis. These results will help prioritise efforts to improve the modelling and calibration of systematic effects in \Euclid.}


   \maketitle
   
\section{Introduction}
Large-scale galaxy surveys are now providing us with very competitive constraints on cosmological models. Observational programmes such as the spectroscopic redshift surveys Sloan Digital Sky Survey \citep[SDSS,][]{Alam_2015} and the Dark Energy Spectroscopic Instrument \citep[DESI,][]{masot2025fullshapeanalysispowerspectrum}, and the photometric surveys Dark Energy Survey \citep[DES,][]{descollaboration2025darkenergysurveyyear}, Kilo Degree Survey \citep[KiDS,][]{KIDSLegacy}, and Hyper Supreme Cam \citep[HSC,][]{HSCY3}, have significantly improved our understanding of the large-scale structure of the universe. Experiments that have recently commenced data collection, such as the \Euclid space telescope~\citep{EuclidSkyOverview} and the Vera C.\ Rubin Observatory’s Legacy Survey of Space and Time \citep[LSST,][]{2009arXiv0912.0201L,2018arXiv180901669T,Ivezic:2008fe} are expected to push these constraints even further. The \Euclid mission stands out as a particularly powerful probe of fundamental physics by combining photometric and spectroscopic probes to provide extremely tight constraints on the time evolution of dark energy and deviations from general relativity at large scales.

It has become increasingly clear that one of the primary challenges in the cosmological analysis of such probes is the size of the parameter space to be explored. Although the concordance \LCDM model only relies on six cosmological parameters, exploring extensions to this model requires larger parameter spaces, often with some of the extra parameters being poorly constrained \citep[see e.g.,][]{Ivanov2020, 2020JCAP...12..018G}. Moreover, each observable is affected by a different set of systematic effects that need to be taken into account. Instrumental calibration, photometric redshift uncertainties, as well as modelling uncertainties in galaxy clustering and weak lensing all introduce biases that must be carefully mitigated to extract unbiased cosmological information \citep{Paykari-EP6, Cragg_2022, Lepori-EP19, Tanidis-TBD, Awan_2025}.

Previous analyses have devoted significant efforts to identifying many systematic effects that are present in large-scale galaxy surveys and understanding how to mitigate their impact \citep{Liu_2016, Berlfein_2024}. In many cases, this is done by modelling the systematic effect with a parametric model and treating its parameters as nuisance parameters, which will be marginalised before presenting the constraints on the cosmological parameters \citep{Myles_2021}. The main problem of allowing these nuisance parameters to vary and absorb the residual systematic effects is that part of the constraining power will be devoted to constraining these parameters, which generally do not contain relevant cosmological information.

One option to compensate for the loss of constraining power is to add informative priors on the nuisance parameters. This is only possible when the choice of the prior can be motivated by our previous knowledge of the performance of the instrument or the physics that generates the systematic uncertainty \citep[see again][]{Myles_2021}. In fact, using ill-motivated priors that might be too tight or incorrectly located might end up biasing the main cosmological parameter constraints and result in false detections of physics beyond the standard cosmological model \citep{Sun_2009}.

Because of all this, it is very important to understand at which level systematic effects can impact the cosmological results and how sensitive our baseline results are to different choices of modelling and priors on the systematic uncertainties. In this article, we explicitly address this point by answering the following questions.
\begin{enumerate}
    \item Can we fix some of the nuisance parameters to reduce the size of the parameter space without incurring biases on the cosmological parameters?
    \item Is our modelling of systematics too complex, resulting in poorly constrained nuisance parameters?
    \item Is our modelling of systematics too simplistic or incorrect, resulting in biases on the cosmological parameters?
    \item Is our prior too conservative, resulting in loss of constraining power?
    \item Is our prior too optimistic, resulting in biases on the cosmological parameters?
\end{enumerate}
The answers to these questions will guide us in how to define an optimal analysis set-up that maximises the cosmological information while ensuring unbiased constraints from \Euclid data. We base our work on the survey specifications for the final \Euclid Data Release and use \CLOE, the official \Euclid likelihood code, which is presented in \cite{Cardone24}, \cite{Joudaki24}, \cite{Canas-Herrera24}, and \cite{Martinelli24}. 

The paper is organised as follows. In Sect.~\ref{sec:sys_effects_photo}, we present the main systematic effects that are present in \Euclid photometric observables. The systematic effects that are important for the spectroscopic observables are described in Sect.~\ref{sec:syst_effects_spectro}. We then present the list of priors and baselines considered in this analysis in Sect.~\ref{sec:methodology} and the main results of our analysis in Sect.~\ref{sec:results}. Finally, we provide a discussion in Sect.~\ref{sec:discussion}.

\section{Photometric observables and their systematic effects}\label{sec:sys_effects_photo}

The imaging survey of \Euclid will provide the 2D positions, photometric redshifts and shapes of almost 1.5 billion galaxies. Based on these measurements, we will be able to construct the 3$\times$2pt summary statistics of the so-called photometric galaxy clustering (GCph), the weak lensing (WL), and the galaxy-galaxy lensing (XC) observable, which is the cross-correlation between galaxy clustering and weak lensing. We will only report here the relevant definitions and refer the reader to \cite{Cardone24} for the full description of the observables and their theoretical modelling.

In this study, we focus on the angular power spectra, which we denote as $\cl{\ell}[ij][AB]$ at the angular multipole $\ell$ for observable $A$ in redshift bin $i$ and observable $B$ in redshift bin $j$. These can be expressed as integrals of the 3D power spectra $P^{\rm photo}_{AB}(k,z)$ at wavevector $k$ and redshift $z$ as
\begin{equation}
\cl{\ell}[ij][AB] = \int{ \de z \,
\frac{c \, W_{i}^{A}(\ell, z) \, W_{j}^{B}(\ell, z)}
{H(z)\, f_{K}^{2}[r(z)]} P^{\rm photo}_{AB}\left [\frac{\ell + 1/2}{f_K[r(z)]}, z \right ] \,}\;,
\label{eq: cijabgen}
\end{equation}
where $W^A_i(\ell,z)$ is the $\ell$- and $z$-dependent kernel, $H(z)$ is the Hubble function and $f_{K}[r(z)]$ is the comoving distance for a curvature value $K$. We use the Limber approximation \citep{Kaiser1992}, which is valid for $\ell\gg1$ and is accurate in the case of broad redshift bins (as are the photometric estimates). Under this limit, we relate the wavevector $k$ to the multipole $\ell$ as
\begin{equation}
    k=\frac{\ell + 1/2}{f_K[r(z)]}\;.
\end{equation}
In this work, we use the \texttt{HMcode} \citep{10.1093/mnras/stab082} non-linear emulator described in Euclid Collaboration: Carrilho et al. in prep. for the 3D power spectra $P^{\rm photo}_{AB}(k,z)$.

\subsection{Weak lensing}
\label{subsec:WL}
It is well known that the intervening matter deflects the paths of photons emitted by distant sources, resulting in the distortion of their images. This distortion can be decomposed into the spin-0 convergence field $\kappa$ and the spin-2 shear field $\gamma$. These describe the size magnification (circles) and the distortion of the galaxy shape (ellipses) in different directions, respectively. Both contain useful cosmological information, but the convergence is harder to model due to the necessity of prior knowledge on the true size of the source \citep{2013MNRAS.433L...6H,Alsing2015} and is also more prone to bias \citep{10.1093/mnras/stx724}. Therefore, the common WL observable in large-scale-structure surveys is cosmic shear.

The correlation of galaxy shapes across different scales thus carries information about the distribution of matter along the line of sight. However, cosmological information is contaminated by the intrinsic correlation of shapes due to local interactions between source galaxies, an effect called intrinsic alignments \citep[hereafter IA,][]{Joachimi2015}. The WL power spectra will then be a combination of shear and IA correlations, which can be expressed as
\begin{equation}
\cl{\ell}[ij][\rm W \rm L] = \cl{\ell}[ij][\rm \gamma \rm \gamma] + \cl{\ell}[ij][\rm I \rm I]+ \cl{\ell}[ij][\rm \gamma \rm I]\;,
    \label{eq: WL}
\end{equation}
where $\cl{\ell}[ij][\rm \gamma \rm \gamma]$ is the shear power spectrum, $\cl{\ell}[ij][\rm \gamma \rm I]$ is the cross-power spectrum between shear and IA and $\cl{\ell}[ij][\rm I \rm I]$ is the IA power spectrum.

The weak lensing observables are also subject to biases due to instrumental effects \citep{2013MNRAS.431.3103C} and systematic errors due to mis-determination of the redshift distribution of the sources and lenses \citep{Desprez-EP10, Naidoo23}. These effects are expected to be kept under control thanks to the data calibration process, and therefore it is considered safe to use Gaussian priors for the nuisance parameters related to them \citep{Cropper16, Martinet-EP4, Schirmer-EP18, Jansen24, EuclidSkyVIS, EP-Congedo, Q1-TP002}.

\subsubsection{Intrinsic alignments}
The IA of galaxy shapes, as we introduced in Sect.~\ref{subsec:WL}, is caused by tidal gravitational fields in large-scale structures that align nearby galaxies, as well as events such as galaxy mergers, which influence the relative alignments of both shapes and angular momenta throughout their evolution. To model this signal, we compute $\cl{\ell}[ij][\rm I \rm I]$ and $\cl{\ell}[ij][\rm \gamma \rm I]$ as in Eqs. (33--35) of \citet{Cardone24}. These involve the $P_{{\rm II}}$ and $P_{{\rm Im}}$ power spectra respectively, which we relate to the matter power spectrum $P_{\rm mm}$ as
\begin{align}
P_{{\rm II}}(k,z)&=[f_{\rm IA}(z)]^2\,P_{{\rm mm}}(k,z)\;,\\
P_{{\rm Im}}(k,z)&=f_{\rm IA}(z)\,P_{{\rm mm}}(k,z)\;.
    \label{eq:IA_power}
\end{align}
The IA prefactor is defined as
\begin{equation}
f_{\rm IA}(z)=-\mathcal A_{\rm IA} \,\mathcal C_{\rm IA} \,\frac{{\ensuremath{\Omega_{{\rm m},0}}}\,}{D\left(z\right)} \left(\frac{1+z}{1+z_0}\right)^{\eta_{\rm IA}}\;,
    \label{eq:in_al}
\end{equation}
with $\Omega_{{\rm m},0}$ the matter fraction today, $D\left(z\right)$ the linear growth factor and $z_0$ the pivot redshift (arbitrarily set to zero here). This is called the redshift-dependent nonlinear alignment (zNLA) model \citep{Joachimi_2011}. The nuisance parameters of the model are $\mathcal A_{\rm IA}$ and $\eta_{\rm IA}$ while the parameter $\mathcal C_{\rm IA}$ is fixed at the value $5\times10^{-14}\,h^{-2}\,M_{\odot}\,{\rm Mpc}^{-3}$ as it is degenerate with $\mathcal A_{\rm IA}$ \citep{EuclidSkyOverview}. While there are more sophisticated IA models in the literature \citep[e.g. TATT,][]{2015JCAP...08..015B}, as we will show in the following, we can already see significant effects due to mis-modelling even within this simpler model.

\subsubsection{Shear bias}
Additional contamination of the shear field can come from instrumental effects such as point spread function (PSF) modelling errors, selection effects, blending, or measurement errors of the galaxy shape \citep{Huterer2006, Massey_2013, Mandelbaum2018}. These contributions can be multiplicative or additive to the shear signal \citep{Massey_2013}. For example, PSF size modelling errors and pixel noise cause a multiplicative effect \citep{Melchior_2012, Refregier_2012,Bernstein_2010}, while the anisotropy of the modelled PSF, induced by elliptical PSFs or unaccounted for charge transfer inefficiencies, can cause an additive effect. The former can be determined using image simulations \citep{10.1093/mnras/stx724, Kannawadi_2019, Pujol2020} and the latter can be inferred from the data itself using empirical null tests \citep{van_Uitert_2016,2023A&A...679A.133L}. Furthermore, both of these effects can be redshift- or scale-dependent. For instance, colour-dependent PSF errors related to an incorrect modelling of the spectral energy distributions (SEDs) of galaxies induce redshift-dependent shear biases \citep{10.1093/mnras/stt602, 10.1093/mnras/sty830, Schutt_2025}, and the thermal variations due to the Solar aspect angle of the spacecraft as it observes in orbit induce spatially varying contributions \citep{Cragg_2022}. Once determined, the shear biases are subtracted from the data. However, we expect some residuals to remain, especially from the redshift-dependent multiplicative bias. Accounting for these residuals amounts to transforming the WL power spectra with a multiplicative factor that depends on the parameter $m_i$ for each redshift bin $i$ as
\begin{equation}
\cl{\ell}[ij][\rm W \rm L] \longrightarrow (1+m_i)\,(1+m_j)\, \cl{\ell}[ij][\rm W \rm L]\;.
    \label{eq: m_bias}
\end{equation}
Although in principle multiplicative bias can be scale dependent, \citet{2019OJAp....2E...5K} showed that it is sufficient to consider a mean factor $m$, provided the spatial variations are small.

\subsubsection{Redshift shifts in WL}
In order to account for biases in the calibration of the photometric redshift distribution, we allow for a shift $\Delta z_i$ in the mean of the redshift distribution of the sources $n_{i}^{\rm S}(z)$ in each bin $i$, following \citet{Abbott_2016}, such that
\begin{equation}
n_{i}^{\rm S}(z) \longrightarrow n_{i}^{\rm S}\,(z-\Delta z_i)\;.
    \label{eq:n_z_shift_sources}
\end{equation}
The reference values of the $\Delta_z$ parameters are derived from the Flagship 2.1 simulation \citep{EuclidSkyFlagship}.

\subsection{Photometric galaxy clustering}
We measure photometric galaxy clustering (GCph) by correlating the 2D positions of galaxies on the sky in tomographic redshift bins and measuring their angular power spectrum $\cl{\ell}[\rm GG]$. This receives contributions from the clustering of galaxies within the sample and from galaxies that scatter in and out of the sample due to lensing magnification, which modify both their size and luminosity. Therefore, the galaxy clustering angular power spectra take the form
\begin{equation}
\cl{\ell}[ij][\rm GG] = \cl{\ell}[ij][\rm gg] + \cl{\ell}[ij][\mu \mu] + 
\cl{\ell}[ij][\rm g\mu]\;,
\label{eq:GC}
\end{equation}
where the first term is the galaxy-galaxy, the second the magnification-magnification and the third the galaxy-magnification contribution.

\subsubsection{Galaxy bias}
Galaxies are biased tracers of the matter density field \citep{Kaiser_1987}. The bias can in principle be a function of both redshift and scale, but in this work we will consider a linear scale-independent bias \citep{Abbott:2018ydy}. The galaxy-galaxy power spectrum then reads
\begin{equation}
P_{\rm gg}^{{\rm photo}}\left [ 
\frac{\ell + 1/2}{f_K[r(z)]},\, z \right ]=b^2(z)\,P_{\rm mm}^{{\rm photo}}\left [ 
\frac{\ell + 1/2}{f_K[r(z)]},\, z \right ]\;,
    \label{eq:galaxy_power_spectrum}
\end{equation}
where $b(z)$ the redshift-dependent galaxy bias. To model the redshift dependence, we measure the linear bias in each of the photometric redshift bins in the Flagship 2.1 simulation \citep{EuclidSkyFlagship}. We then fit a polynomial function,
\begin{equation}
    b(z) = \sum_{i=0}^3 b_{{\rm gph},i}\,z^i
    \label{eq:galbias_pol_fit}\;,
\end{equation}
and use the $b_{{\rm gph},i}$ parameters to build our reference data vector. In the following, we will also explore the case where the bias is modelled as a constant within each tomographic redshift bin $i$, where the reference values are those obtained from the Flagship 2.1 simulation \citep{EuclidSkyFlagship}.

\subsubsection{Linear redshift-space distortion (RSD)}
The linear RSD signal results in the squashing of the galaxy power spectrum at large scales in the direction perpendicular to the line of sight due to the coherent motion of galaxies towards large and highly dense regions \citep{Kaiser_1987}. In harmonic space and in the Limber approximation, it takes the form as described by the Eqs. (45-47) of \cite{Cardone24}. We note that in this formalism we do not introduce extra parameters to marginalise over, but rather a fixed correcting term per redshift bin to add in the kernel for galaxy clustering. The physical meaning of these terms is that, effectively, at large scales, galaxies are shuffled among neighbouring redshift bins \citep{Tanidis_2019}. Since the significance of the linear RSD in the photometric observables for \Euclid to avoid biases (up to $\sim5\,\sigma$) on the cosmological parameters has been studied thoroughly in \cite{Tanidis-TBD}, we do not test their impact here. However, we always include them in the computation of the data and the theory vectors.

\subsubsection{Magnification bias}
The second and third terms of \cref{eq:GC} are shown in Eqs.\ (42-43) of \cite{Cardone24}. The magnification kernel $W_{i}^{\mu}(z)$ includes the function $b_{\rm mag}(z)$, which is the magnification bias \citep{Schmidt2009a, Schmidt2009b}. This is a lensing contribution in the galaxy clustering of the lenses sample related to the size magnification of the galaxies. It is defined as $b_{\rm mag}(z) = 5 \,s(z) - 2$, where $s(z)$ is the logarithmic slope of the luminosity function for the lenses \citep{Scranton_2005,Menard2010,Hildebrandt}. The magnification effect is two-fold. On the one hand, it means that the distribution of the objects in the images is over larger angles, thus reducing their number density; on the other hand, the same effect allows for the observation of fainter objects. The impact of ignoring the magnification bias in galaxy surveys has been shown in \cite{10.1093/mnras/stt2060} and \cite{Tanidis_2019b}. The effect of neglecting it in the \Euclid photometric observables has been studied in \cite{Lepori-EP19}, where they measured biases up to $\sim 6\,\sigma$ on the cosmological parameters. Similarly to the galaxy bias, we consider two models for the magnification bias. The first is to account for the magnification bias with one parameter per tomographic redshift bin (evaluated at the centre of the bin), and the second is a polynomial function of redshift,
\begin{equation}
    b_{\rm mag}(z) = \sum_{i=0}^3 b_{{\rm mag},i}\,z^i\;,
        \label{eq:magbias_pol_fit}
\end{equation}
which is expected to be a good approximation given that the redshift evolution of the magnification bias is smooth. We define our reference values for the magnification bias parameters using the measurements from the Flagship 2.1 simulation \citep{EuclidSkyFlagship}.

\subsubsection{Redshift shifts in GC}
As we did for the source sample, we also account for biases in the photometric redshift distribution of the lens sample $n_{i}^{\rm L}(z)$ as
\begin{equation}
n_{i}^{\rm L^\prime}(z)=n_{i}^{\rm L}\,(z-\Delta z_i)\;.
    \label{eq:n_z_shift_lenses}
\end{equation}
The reference values of the $\Delta_z$ parameters are derived from the Flagship 2.1 simulation \citep{EuclidSkyFlagship}.

\subsection{Galaxy-galaxy lensing}
It has been shown (e.g. \citealt{JK12, Isaac2020}) that the cross-correlation spectra between the weak lensing and the galaxy clustering observables (also called galaxy-galaxy lensing) significantly improve the constraining power on the cosmological parameters. Therefore, we also include them in this analysis and define them as
\begin{equation}
\cl{\ell}[ij][\rm GL] = 
\cl{\ell}[ij][\rm g\gamma] +
\cl{\ell}[ij][\rm gI] + 
\cl{\ell}[ij][\rm \mu \gamma] + 
\cl{\ell}[ij][\rm \mu I] \ , 
\label{eq: cij_weak_lens_gal_cl}
\end{equation}
with the individual terms described in Eqs. (55--57) of \cite{Cardone24}. The terms on the right-hand side of \cref{eq: cij_weak_lens_gal_cl} represent the cross-correlation between galaxy positions and cosmic shear, galaxy positions and intrinsic alignment, magnification and cosmic shear, and magnification and intrinsic alignment, respectively.  These terms do not introduce any additional systematic effects.

\section{Spectroscopic observables and their systematic effects}\label{sec:syst_effects_spectro}
The spectroscopic survey of \Euclid will target emission-line galaxies and measure their redshift using the $\Halpha$ emission line for around 30 million galaxies, making it possible to map their 3D distribution \citep{EuclidSkyOverview}. The main cosmological probe will be the anisotropic galaxy power spectrum expanded in Legendre multipoles $\pl{k}{\ell}[z]$, as defined in Eq. (90) of \cite{Cardone24}. We use the nonlinear model described in Euclid Collaboration: Crocce et al. in prep., which is based on the Effective-Field Theory of Large-Scale Structure (EFTofLSS, see e.g. Euclid Collaboration: Moretti et al. in prep. and references therein). This model comes with a series of nuisance parameters that are known to cause issues like prior-volume and projection effects \citep{2023PhRvD.107l3530S,2023JCAP...01..028C,2023JCAP...06..005M,2023MNRAS.526.3461D,2024MNRAS.532..783Z,2023PhRvD.108l3514H,2025arXiv250811811M}. The strategy to mitigate such effects has been investigated in Euclid Collaboration: Pezzotta et al. in prep. and is therefore beyond the scope of this work.

\subsection{Redshift errors}
The uncertainty in the spectroscopic redshift estimation causes a smearing of the distribution of galaxies along the line of sight, which must be taken into account in the model. This is done by multiplying the model for the power spectrum by a factor
\begin{equation}
F_z(k, \mu_k, z) = {\rm e}^{-k^2 \,\mu_{k}^{2}\, \sigma_{r}^{2}(z)}\;,
\end{equation}
where $\mu_k$ is the cosine of the angle between $\vec{k}$ and the line of sight, and $\sigma_{r}(z)$ is the error in the estimation of the comoving distance, which is related to the redshift error $\sigma_{z}$ through
\begin{equation}
\sigma_r(z) = \frac{c \,\sigma_{\rm z}}{H(z)}\;.
\end{equation}
In this work, we fix $\sigma_z=0.002$, which is a conservative estimate of the expected precision of the spectroscopic redshift estimation (Euclid Collaboration: Grannet et al. in prep.).

\subsection{Purity and completeness}
The sample of \Halpha galaxies can be contaminated by galaxies whose redshift is misestimated due to line misidentification (line interlopers) or random redshift errors (noise interlopers). These galaxies will impact the clustering signal in distinct ways: line interlopers will have their own clustering signal which will be mixed up with the clustering of \Halpha galaxies, while noise interlopers will lead to a fraction of galaxies in the sample being random outliers that do not cluster. Both of these effects result in a damping of the clustering signal, with an additional deterministic scale-dependence in the case of line interlopers alone.

The number density of measured objects can thus be related to the correct one as
\begin{equation}
    n_{\rm meas}(z)=n_{\rm corr}(z)\,\frac{1-f_{\rm inc}}{1-f_{\rm out}}\;,
\end{equation}
where $f_{\rm inc}=1-f_{\rm comp}$ is the fraction of undetected \Halpha galaxies, $f_{\rm comp}$ is the fraction of correctly identified \Halpha galaxies with respect to the total \Halpha population (completeness), $f_{\rm out}=1-f_{\rm pur}$ is the fraction of outliers, and $f_{\rm pur}$ is the fraction of actual \Halpha galaxies in the sample (purity).

In this work, we only model the effect of noise interlopers on the galaxy power spectrum and leave the modelling of line interlopers to future work. Preliminary works, such as \citet{EP-Risso} and Euclid Collaboration: Lee et al. in prep., showed that ignoring the effect of line interlopers does not bias the cosmological parameter constraints, so we do not expect this choice to have a significant impact on the results presented here. The measured galaxy power spectrum multipoles will then become
\begin{equation}
    \pl{k}{\ell, {\rm meas}}[z] = (1-f_{\rm out})^2\,\pl{k}{\ell, {\rm corr}}[z]\;.
\end{equation}
We consider $f_{\rm out}$ as a nuisance parameter and model it as a constant within each redshift bin. Our reference values are motivated by end-to-end simulations that include interlopers (Euclid Collaboration: Grannet et al. in prep.), and match the ones used in \citet{EuclidSkyOverview} and \citet{Canas-Herrera24}.

\section{Method}
\label{sec:methodology}
The aim of this work is to quantify the impact of different modelling and prior choices on the best-fit values and parameter errors, with a focus on cosmological parameters. We start by defining a synthetic data vector and a set of modelling choices, which are then used to compute the likelihood with \CLOE. Given a choice for the prior on the parameters, we then sample the posterior distribution using the \nautilus nested sampling algorithm \citep{nautilus}. Following the same approach as \cite{Canas-Herrera24}, we generate noiseless synthetic data vectors with \CLOE itself, thus ensuring that any parameter shifts or changes in the constraining power that we observe in our results only arise from the analysis choices that we are testing. For all the tests performed here, we only change the data or the theory vectors and use the data covariance of the baseline setup. A detailed description of the computation of covariances can be found in Section~5 of \cite{Canas-Herrera24}. Since these are analytical covariances, we assume that the likelihood is Gaussian.

Our baseline setup is the same as the one described in \citet{Canas-Herrera24}, which forecasts the expected performance and sample sizes of the final \Euclid Data Release. For the photometric observables, we assume a magnitude-limited sample of galaxies with $\IE \leq 24.5$ distributed in 13 equi-populated photometric redshift bins between redshift 0.2 and 2.5 \citep{EuclidSkyOverview}. The number density of galaxies was derived using the Flagship 2.1 simulation \citep{EuclidSkyFlagship}. In our analysis, we consider the high scale cut scenario, where $\ell_{\rm min} (\text{WL})=\ell_{\rm min} (\text{XC})=\ell_{\rm min} (\text{GCph})=10$ and $\ell_{\rm max}(\text{WL})=5000$,  while $\ell_{\rm max}(\text{GCph})=\ell_{\rm max}(\text{XC})=3000$. For the spectroscopic observables, we consider a flux-limited sample of \Halpha galaxies with $f_{\Halpha}>2\times10^{-16}\textrm{erg}\,\textrm{cm}^{-2}\,\textrm{s}^{-1}$ in 4 redshift bins between redshifts 0.9 and 1.8, centred at $\bar{z}=\{1.0, 1.2, 1.4, 1.65\}$. In this work we analyse the photometric and spectroscopic observables separately. Depending on the analysis choices, the size of the parameter space varies between 16 and 51 dimensions.

In \cref{tab:nuis_fid}, we list the reference values and priors of the nuisance parameters that we study in this work, in the baseline setup. On top of these parameters, we also marginalise over the baryonic feedback efficiency parameter of the \texttt{HMcode} emulator $\log_{10} (T_{\rm AGN}/{\rm K})$, which is centred at 7.75 with a Gaussian width of 0.17825. We do not show the impact of modifying this parameter or its degeneracy with the rest of the nuisance parameters, as this has already been investigated in Euclid Collaboration: Carrilho et al. in prep.

\begin{table}[htbp]
  \centering
  \caption{Reference values and priors on the nuisance parameters we are testing. When the prior is uniform, we specify the range; when Gaussian, we specify the width.}
  \label{tab:nuis_fid}
  \begin{tabular}{@{}lrlc@{}}
    \toprule
    \textbf{Name}      & \textbf{Value}\phantom{aaa}   & \textbf{Prior type} & \textbf{Prior width or range} \\
    \midrule
    \midrule
    \multicolumn{4}{c}{\textbf{Intrinsic alignment parameters}} \\
    \midrule
    $\mathcal{A}_{\rm IA}$       & 0.16\phantom{0000}             & Uniform              & $[-2,2]$                     \\
    $\eta_{\rm IA}$    & 1.66\phantom{0000}             & Uniform              & $[0,3]$                     \\
    \midrule
    \multicolumn{4}{c}{\textbf{Photometric galaxy bias parameters}} \\
    \midrule
    $b_{\rm gph,0}$    & 1.33291\phantom{0}          & Uniform              & $[-3,3]$                     \\
    $b_{\rm gph,1}$    & $-0.72414$\phantom{0}       & Uniform              & $[-3,3]$                     \\
    $b_{\rm gph,2}$    & 1.01830\phantom{0}         & Uniform              & $[-3,3]$                     \\
    $b_{\rm gph,3}$    & $-0.14913$\phantom{0}       & Uniform              & $[-3,3]$                     \\
    \midrule
    \multicolumn{4}{c}{\textbf{Photometric magnification bias parameters}} \\
    \midrule
    $b_{\rm mag,0}$    & $-1.50685$\phantom{0}       & Uniform              & $[-3,3]$                     \\
    $b_{\rm mag,1}$    & 1.35034\phantom{0}          & Uniform              & $[-3,3]$                     \\
    $b_{\rm mag,2}$    & 0.08321\phantom{0}          & Uniform              & $[-3,3]$                     \\
    $b_{\rm mag,3}$    & 0.04279\phantom{0}          & Uniform              & $[-3,3]$                     \\
    \midrule
    \multicolumn{4}{c}{\textbf{Photometric redshift shift parameters}} \\
    \midrule
    $\Delta z_{1}$     & $-0.025749$      & Gaussian             & 0.00257890                    \\
    $\Delta z_{2}$     & 0.022716         & Gaussian             & 0.00275258                   \\
    $\Delta z_{3}$     & $-0.026032$      & Gaussian             & 0.00287484                   \\
    $\Delta z_{4}$     & 0.012594         & Gaussian             & 0.00307264                   \\
    $\Delta z_{5}$     & 0.019285         & Gaussian             & 0.00323726                   \\
    $\Delta z_{6}$     & 0.008326         & Gaussian             & 0.00341852                   \\
    $\Delta z_{7}$     & 0.038207         & Gaussian             & 0.00360394                   \\
    $\Delta z_{8}$     & 0.002732         & Gaussian             & 0.00371824                   \\
    $\Delta z_{9}$     & 0.034066         & Gaussian             & 0.00395192                   \\
    $\Delta z_{10}$    & 0.049479         & Gaussian             & 0.00418664                   \\
    $\Delta z_{11}$    & 0.066490         & Gaussian             & 0.00449286                   \\
    $\Delta z_{12}$    & 0.000815         & Gaussian             & 0.00497880                   \\
    $\Delta z_{13}$    & 0.049070         & Gaussian             & 0.00584318                   \\
    \midrule
    \multicolumn{4}{c}{\textbf{Multiplicative bias parameters}} \\
    \midrule
    $m_{i\in\{1,13\}}$            & 0\phantom{.000000}                & Gaussian             & 0.002\phantom{00000}                        \\
    \midrule
    \multicolumn{4}{c}{\textbf{Outlier fraction parameters}} \\
    \midrule
    $f_{\rm out,1}$    &        0.195\phantom{000}          &  Gaussian                      &                  0.01\phantom{000000}            \\
    $f_{\rm out,2}$    &        0.204\phantom{000}        &    Gaussian                   &                    0.01\phantom{000000}          \\
    $f_{\rm out,3}$    &        0.306\phantom{000}          &  Gaussian                     &                  0.01\phantom{000000}            \\
    $f_{\rm out,4}$    &        0.121\phantom{000}          &  Gaussian                      &                  0.01\phantom{000000}            \\
    \bottomrule
  \end{tabular}
\end{table}

\subsection{Quantification of biases and constraining power}
When assessing the shifts in a parameter $\theta$ with respect to the baseline model, we compute the following quantity:
\begin{equation}
    {\Delta\sigma}_{\theta}=\frac{\left|\,\overline{\theta}_{\rm base}-\overline{\theta}\,\right|}{\sqrt{\sigma_{\theta, {\rm base}}^2+{\sigma_{\rm{\theta}}^2}}}\;,
    \label{eq:shifts}
\end{equation}
where $\overline{\theta}_{\rm base}$ and $\overline{\theta}$ are the mean values of the parameter posterior in the baseline model and in the model we want to examine, while $\sigma_{\theta, {\rm base}}$ and $\sigma_{\theta}$ are their corresponding $1\,\sigma$ uncertainties. Equation~\eqref{eq:shifts} is equivalent to the figure of bias under the assumption that the posteriors are Gaussian and uncorrelated. We notice that this is a conservative choice, since choosing a metric that takes the correlations into consideration would lead to larger values for the parameter shifts.

We also examine cases where we keep the data vectors unchanged and only test the impact of changing the prior width of some parameters. This is quantified in terms of constraining power, which is simply computed as the percent difference between the parameter errors with the baseline prior width and with the different width that we are testing.

\section{Results}\label{sec:results}
In this section we present the results of our analysis in terms of constraints on the cosmological parameters. Figures only show parameter contours that display significant changes with respect to the baseline setup (always in black). Relevant subsets of nuisance parameter constraints are shown in Appendix~\ref{sec:appendix}. 

The common practice to test for robustness of the analysis pipeline is to check that bias due to different choices in the modelling stays within a certain fraction of the statistical uncertainty. In the following, we will report all cases where ${\Delta\sigma}_{\theta}>0.1$, i.e. more than 10\% of the statistical uncertainty. 

A summary of the shift in the best-fit parameters and change in the constraining power is given in Tables \ref{tab:param-shifts} and \ref{tab:merged-constraint-shifts}, respectively. All the files containing the samples for each case can be found at the following link.\footnote{\href{https://doi.org/10.5281/zenodo.16029827}{https://doi.org/10.5281/zenodo.16029827}}

\begin{table}[htbp]
  \centering
  \caption{Parameter shifts under various data/model assumptions. Every value in this table is the parameter shift ${\Delta\sigma}_{\theta}$ defined in \cref{eq:shifts} for the model we test compared to the baseline model. For each case, we specify the data (d), the model (m), or the prior (p), where $\mathcal{G}(m,\sigma)$ indicates a Gaussian centred in $m$ with width $\sigma$. We note that we report changes only on the cosmological parameters which are of interest for each model. Unless otherwise stated in parentheses, it is always a 3$\times$2pt analysis.}
  \label{tab:param-shifts}
  \begin{tabularx}{\columnwidth}{@{}lcccccc@{}}
    \toprule
    \multicolumn{7}{@{}c}{\textbf{Intrinsic alignment}} \\
    \midrule
     & $H_0$ & $\Ob$ & $\Om$ & $\ns$ & $\sotto$ & $S_8$ \\
    \midrule
    d: no IA, m: zNLA & 6.54 & 4.66 & 5.48 & 5.69 & 4.82 & 2.72 \\
    d: zNLA, m: no IA & 0.27 & 0.41 & 1.35 & 0.26 & 0.06 & 2.29 \\
    \rotatebox[origin=c]{180}{$\Lsh$} (2$\times$2pt) & 0.14 & 0.32 & 0.68 & 0.38 & 0.66 & 2.11 \\
    \rotatebox[origin=c]{180}{$\Lsh$} (WL) & 0.61 & 2.04 & 4.23 & 1.08 & 4.49 & 1.00 \\
    \midrule
  \end{tabularx}
  \begin{tabularx}{\columnwidth}{@{}lcccc@{}}
    
    \multicolumn{5}{@{}c}{\textbf{Galaxy bias}} \\
    \midrule
     & $\Om$ & $\ns$ & $\sotto$ & $S_8$ \\
    \midrule
    d: polynomial, m: const. per-bin & 0.25 & 0.42 & 1.28 & 1.53 \\
    d: polynomial, m: interp. per-bin & 0.29 & 0.11 & 0.56 & 0.30 \\
    \midrule
    \multicolumn{5}{@{}c}{\textbf{Multiplicative bias}} \\
    \midrule
     & $\Om$ & $\ns$ & $\sotto$ & $S_8$ \\
    \midrule
    d: $m_i=0.002$, p: $\mathcal{G}(0,0.002)$ & 0.37 & 0.23 & 0.09 & 0.72 \\
    d: $m_i=0.002$, p: $\mathcal{G}(0,0.004)$ & 0.35 & 0.24 & 0.08 & 0.63 \\
    d: $m_i=0.002$, m: no m-bias & 0.37 & 0.24 & 0.10 & 0.78 \\
    \bottomrule
  \end{tabularx}
\end{table}

\begin{table}[htbp]
  \centering
  \caption{Percentage change in the constraining power of the cosmological parameters for different prior choices. Each value in this table is the percentage difference ($+$ denotes increase and $-$ decrease) between the model parameter we test against the baseline model. Baselines: prior width $\Delta z = 2\times10^{-3}(1+z_i)$ for the redshift shifts, and 1\% prior for the purity parameters.}
  \label{tab:merged-constraint-shifts}
    \begin{tabularx}{\columnwidth}{@{}lrrr@{}}
    \toprule
    \multicolumn{4}{c}{\textbf{Redshift shifts}} \\
    \midrule
     & $\Om$ & $\sotto$ & $S_8$ \\
    \midrule
    $\Delta z_i = 2\times10^{-2}(1+z_i)$ & $-7.32\%$ & $-9.81\%$ & $-0.53\%$ \\
    Correlated $\Delta z=0.0036$ & $-0.09\%$ & $-0.05\%$ & $+0.26\%$ \\
    Fixed to reference & $+11.69\%$ & $+17.38\%$ & $+3.97\%$ \\
    \midrule
  \end{tabularx}
  \textbf{Purity}
  \begin{tabularx}{\columnwidth}{@{}lXrr@{}}
    \midrule
     & & $\lnAs$ & $\Oc h^{2}$ \\
    \midrule
    Prior width: 5\% & & $-48.52\%$ & $-6.68\%$ \\
    Prior width: 10\% & & $-70.23\%$ & $-13.87\%$ \\
    \bottomrule
  \end{tabularx}
\end{table}

\subsection{Intrinsic alignment}
The amplitude of intrinsic alignment \aIA is consistent with 0 for all Stage-III weak lensing surveys \citep{DESY3Secco,HSCY3,KIDSLegacy}. While it is possible that the signal is very small \citep{2021MNRAS.501.2983F} and that with the reduced statistical errors from Stage-IV surveys we will be able to measure it, two scenarios are possible: 
\begin{enumerate}
    \item if the signal is still small compared to the statistical error and we try to model it with a model that is too complex, we could be impacted by prior volume effects and risk biasing the cosmological parameters;
    \item if the signal has a complex dependency on redshift and other galaxy properties and we try to model it with a simplistic model, due to the large amount of parameters that we are varying this mis-modelling could be re-absorbed by other parameters.
\end{enumerate}

\begin{figure}
    \centering
    \includegraphics[width=1\linewidth]{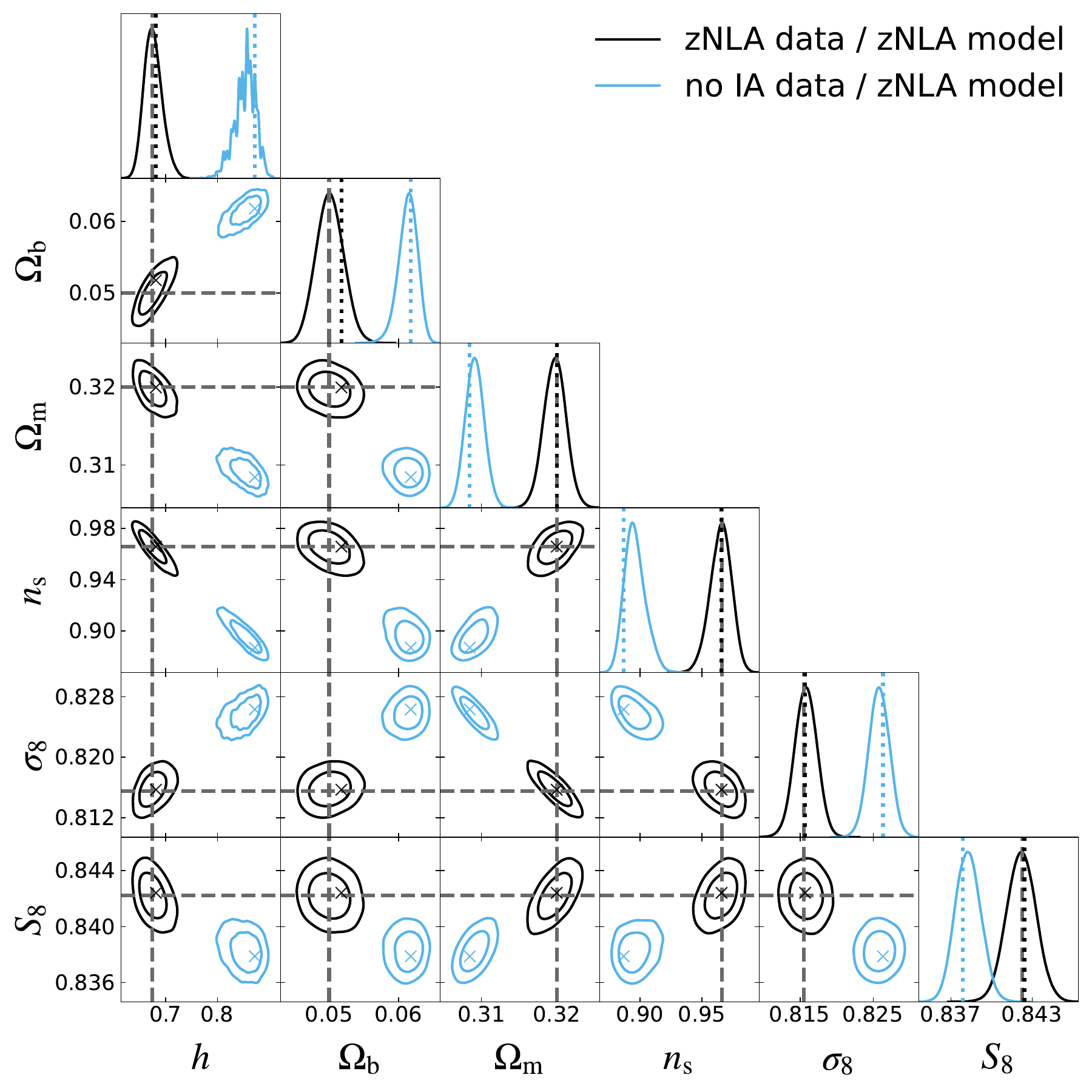}
    \caption{Cosmological parameter constraints in the baseline case (black), and using the zNLA intrinsic alignment model for a data vector with zero intrinsic alignment signal (blue). The MAP value is shown in the respective colours for the 2D (crosses) and 1D (dotted lines) projected posterior panels. \textit{Note:} the 1D projected posterior for $h$ looks scattered because the samples are hitting the upper boundary of the prior.}
    \label{fig:noIA_zNLA_cosmo}
\end{figure}

To address the first case, we analyse a data vector with zero intrinsic alignment signal using the baseline zNLA model. As can be seen in \cref{fig:noIA_zNLA_cosmo}, all of the cosmological parameters show important shifts from the baseline, ranging from 2.72 $\sigma$ for $S_8$ to 6.54 $\sigma$ for $H_0$ (see \cref{tab:param-shifts}). In \cref{fig:noIA_zNLA_nuis}, we show the constraints on the IA parameters and the redshift shifts: as expected, the amplitude \aIA gets close to 0 and the power-law index \etaIA becomes unconstrained, resulting in important shifts on the $\Delta z$ parameters, confirming previous findings in this direction \citep{2023JCAP...01..033F}. Such large shifts can be attributed to prior-volume effects caused by this nested zNLA parameter structure. We show the Maximum A Posteriori (MAP) value in \cref{fig:noIA_zNLA_cosmo} and \cref{fig:noIA_zNLA_nuis} as crosses and dotted lines in the 2D and 1D projected posteriors, respectively. Although in this case the MAP is only marginally different from the projected posterior, we cannot conclude that the prior volume effects are not the root cause of the shifts that we observe. In fact, MAP values might still be affected by prior choices \citep{2022PhRvD.106f3506G,2023PhRvD.108l3514H,2025PhRvD.111h3504H}.

\begin{figure}
    \centering
    \includegraphics[width=1\linewidth]{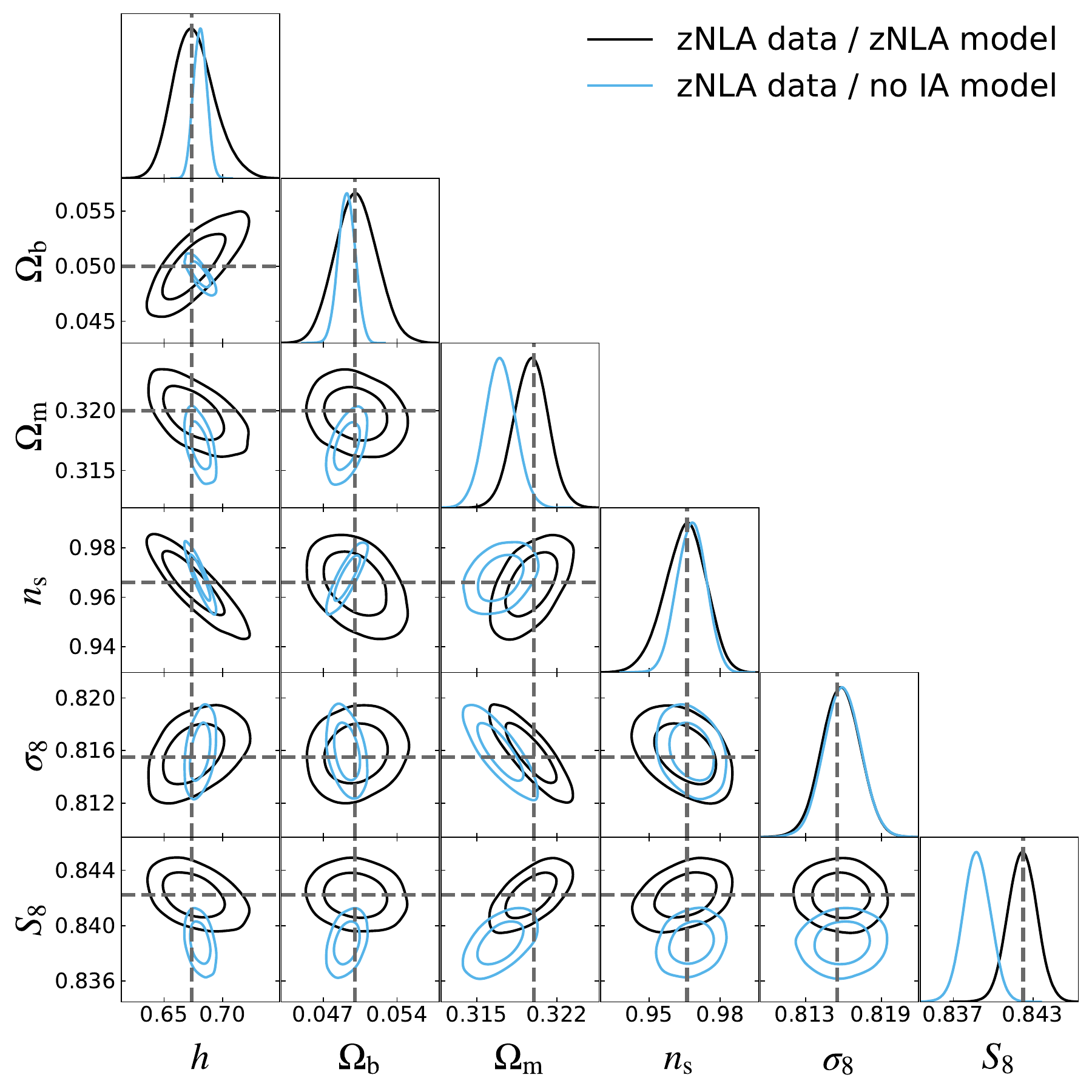}
    \caption{Cosmological parameter constraints in the baseline case (black), and analysing a data vector with zNLA signal with no intrinsic alignment model (blue).}
    \label{fig:IA_mismodelling_cosmo}
\end{figure}

We address the impact of IA mis-modelling by fitting the baseline data vector, which contains a zNLA signal, without modelling IA at all. The results are shown in \cref{fig:IA_mismodelling_cosmo} for the cosmological parameters and in \cref{fig:IA_mismodelling_nuis} for the nuisance parameters. We can observe a bias greater than $0.3\,\sigma$ in most of the cosmological parameters, especially on \Om, \sotto, and $S_8$. We can also see that the size of the contours shrinks because of the smaller parameter space. In terms of nuisance parameters, we can again observe shifts in the $\Delta z$ parameters, albeit smaller than in the previous case.

\begin{figure}
    \centering
    \includegraphics[width=1\linewidth]{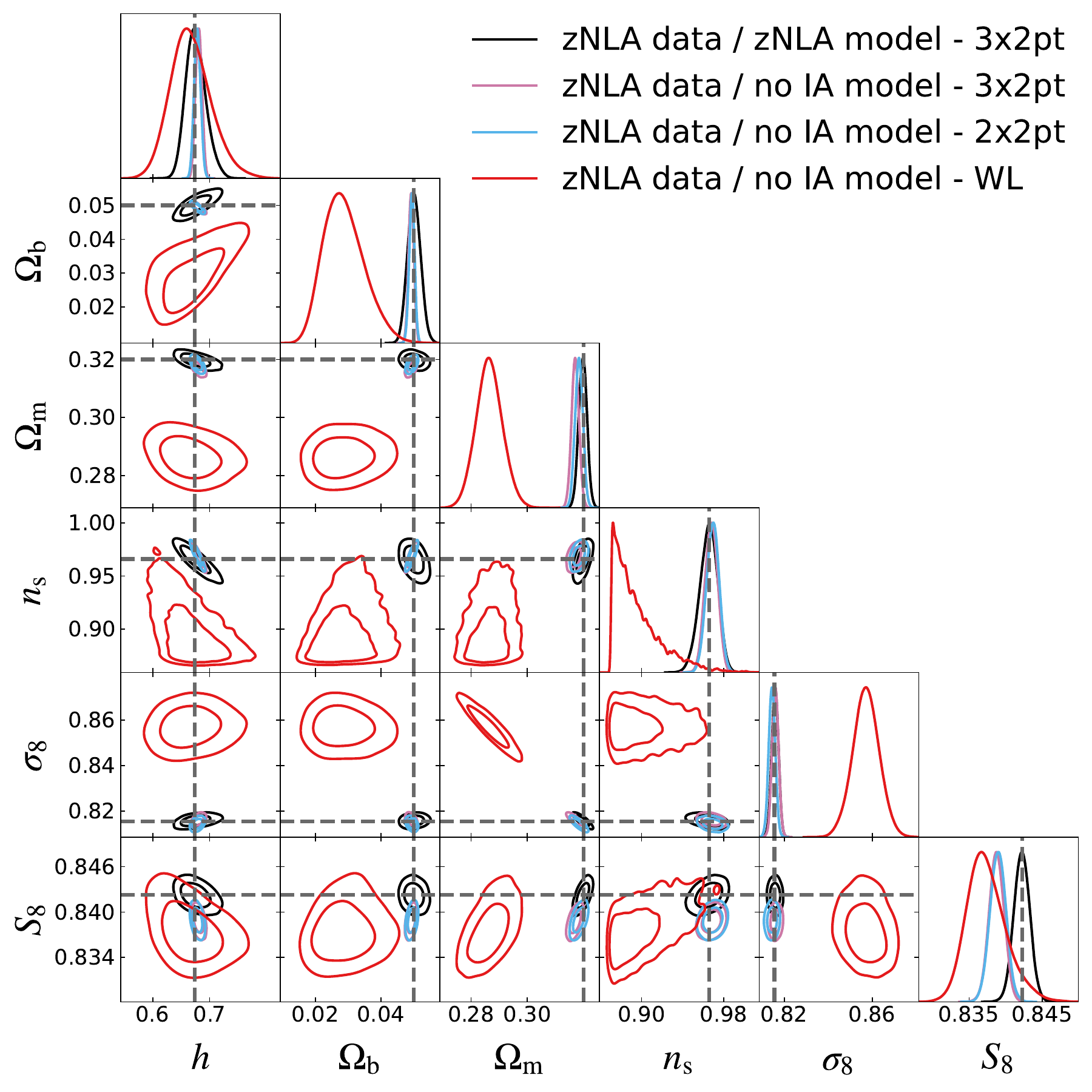}
    \caption{Cosmological parameter constraints in the baseline case (black), and analysing a data vector with zNLA signal with no intrinsic alignment model using the $3 \times 2$ pt combination (purple), $2 \times 2$ pt (blue) and WL only (red).}
    \label{fig:IA_mismodeling_split_cosmo}
\end{figure}

\citet{Samuroff2024} suggest that IA mis-modelling can be more prominently identified when analysing $2\times$2pt and WL separately, where the latter would be significantly more biased. We test this in our \Euclid-specific setup by running a case where we fit the baseline data vector without any IA modelling. We show the results in Figs.~\ref{fig:IA_mismodeling_split_cosmo} and \ref{fig:IA_mismodeling_split_nuis} for the cosmological and nuisance parameters, respectively. We can indeed confirm that, also in our case, WL is significantly more impacted by IA mis-modelling. In fact, we can observe significant biases in the cosmological and $\Delta z$ parameters, with a maximum of $4.49\,\sigma$ for \sotto.

In summary, correctly modelling IA is of utmost importance in recovering unbiased cosmological parameter constraints from a 3$\times$2pt analysis. Some hints that special care is needed are the presence of poorly constrained nuisance parameters and inconsistencies in the cosmological parameters between the different observables. More in-depth tests of IA mis-modelling have been conducted in Euclid -Collaboration: Navarro-Gironés et al. in prep. in the context of the first \Euclid Data Release. Despite having larger statistical errors, their work also shows that IA mis-modelling may induce significant biases in the cosmological parameters.

\subsection{Photometric galaxy bias}
In our baseline setup we choose a polynomial functional form to describe the redshift evolution of the linear galaxy bias for the photometric sample. While this is a sensible choice, since we expect galaxy bias to evolve smoothly with redshift, a more conservative approach would be to use one galaxy bias parameter per redshift bin. This implies fitting 13 parameters instead of 4, significantly increasing the size of the parameter space and the risk of incurring projection effects. 

\begin{figure}
    \centering
    \includegraphics[width=\linewidth]{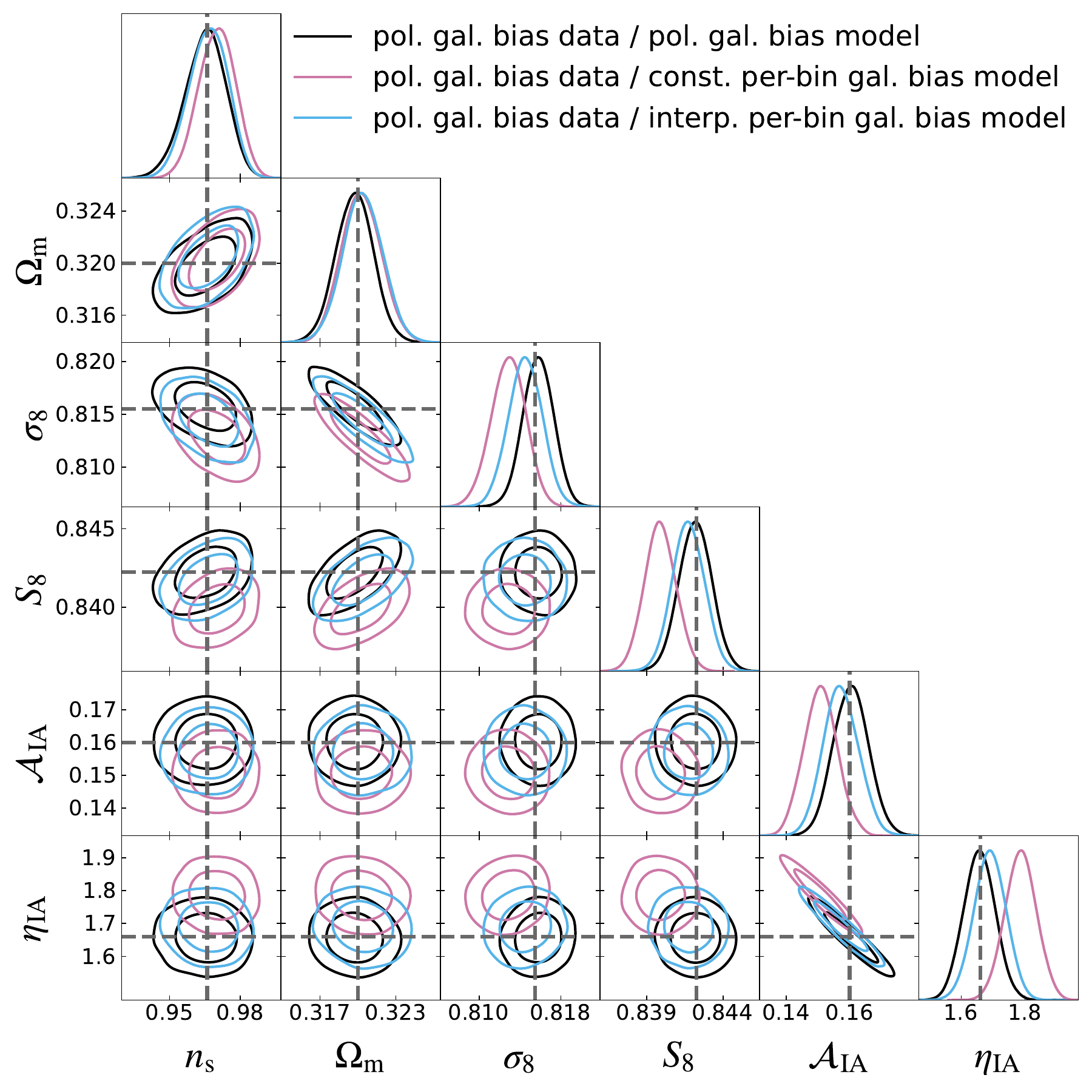}
    \caption{Constraints on cosmological and IA parameters in the baseline case (black), and analysing the baseline data vector with a constant per-bin galaxy bias model (purple) or with a linear function of redshift within each bin (blue).}
    \label{fig:poly_gal_bias_data_per_bin_model_cosmo_params}
\end{figure}

We test this by fitting the baseline data vector, generated with a polynomial galaxy bias function, with a model that has a constant value of galaxy bias within each redshift bin. The results are shown in \cref{fig:poly_gal_bias_data_per_bin_model_cosmo_params} for cosmological and IA parameters, \cref{fig:poly_gal_bias_data_per_bin_model_b_params} for galaxy bias parameters, and \cref{fig:poly_gal_bias_data_per_bin_model_b_params_effect_on_photoz} for the redshift shift parameters. From these figures, we see that using constant per-bin values (in purple) leads to biases on some of the cosmological parameters, with a maximum of $1.53 \,\sigma$ for $S_8$, and much more important biases on all of the nuisance parameters. We have tested that this does not arise from projection effects alone in the optimistic scale cuts employed in this work by repeating the analysis with a cut at $\ell_{\rm max}=1500$ for the GCph and XC spectra, which also resulted in a $1.17\,\sigma$ bias on $S_8$.

Let us note here that the baseline values for the polynomial galaxy bias parameters have been obtained by fitting the values of galaxy bias at the centre of each redshift bin as measured in the Flagship 2.1 simulation \citep{EuclidSkyFlagship}. In the simulation, galaxy bias is evolving within the redshift bin, and our measurement is a weighted average over the redshift range of the bin. Assuming that the bias is constant within the bin leads to an unphysical step-like redshift evolution, which can cause most of the differences seen here. We further check this by fitting the polynomial bias data with the constant per-bin model, but now throwing away the highest redshift bin. This is the widest bin within which there is a lot of redshift evolution that is neglected. We still find a biased constraint (due to the unaccounted-for redshift evolution in the remaining tomographic bins) but somewhat alleviated, finding a shift of $0.98\,\sigma$ for $S_8$. We thus run an additional case where the galaxy bias is assumed to be a linear function within each redshift bin, which is shown in blue in Figs.~\ref{fig:poly_gal_bias_data_per_bin_model_cosmo_params} and \ref{fig:poly_gal_bias_data_per_bin_model_b_params_effect_on_photoz}. We note that this model significantly reduces the biases in all of the parameters, with a maximum shift of $0.56 \,\sigma$ for \sotto.

These tests show that \Euclid data are rather sensitive to the modelling of the redshift evolution of galaxy bias, highlighting the need for a physically motivated modelling of this systematic effect.

\subsection{Magnification bias}
\begin{figure}
    \centering
    \includegraphics[width=\linewidth]{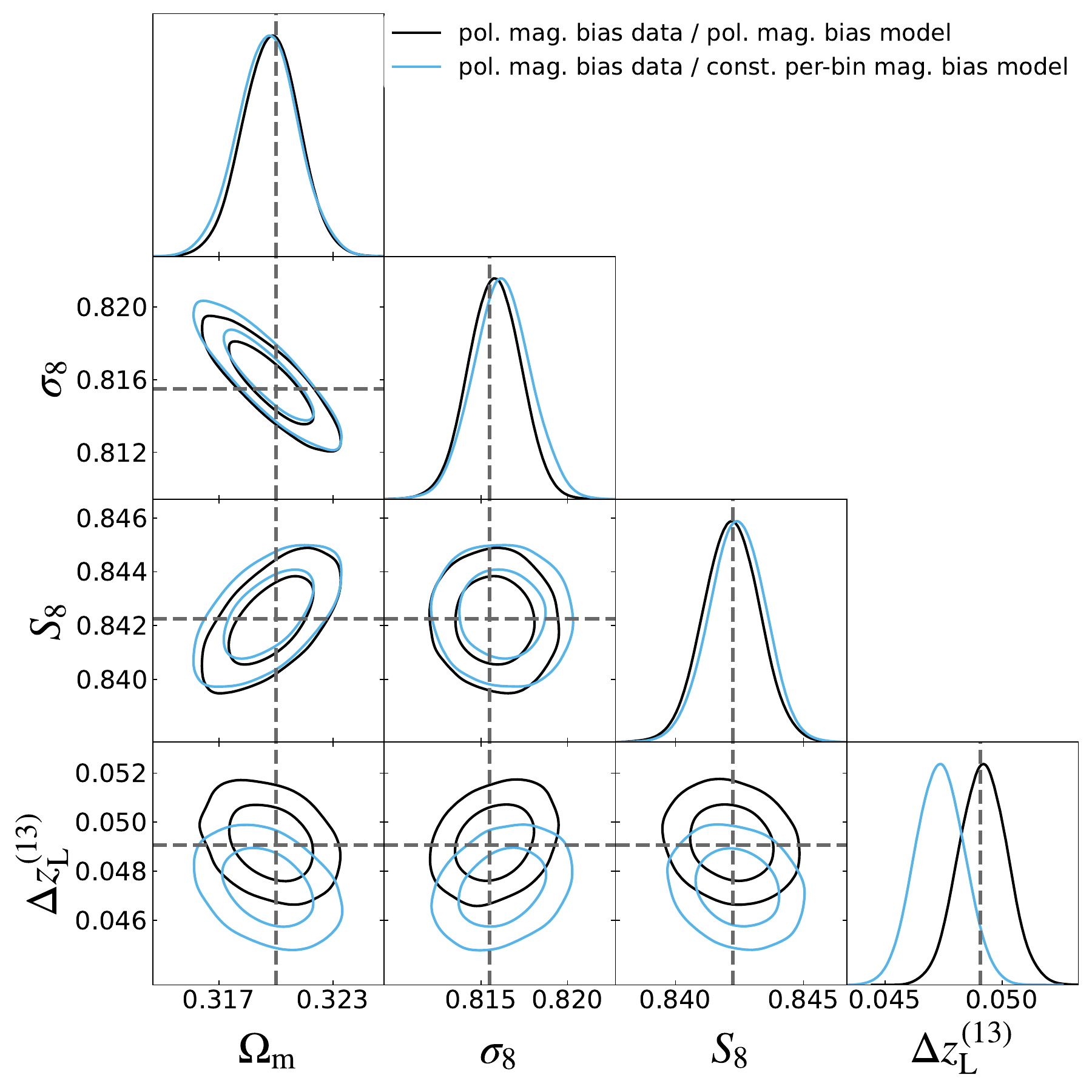}
    \caption{Parameter constraints in the baseline setup (black) and when fitting the baseline data vector with a constant per-bin magnification bias model (blue).}
    \label{fig:mag_bias_per_bin_prior_subset}
\end{figure}

Similarly to the case of galaxy bias, our baseline setup also uses a polynomial function to model the redshift evolution of magnification bias. We test the impact of using a constant per-bin model to fit our baseline data vector and, as we can see in \cref{fig:mag_bias_per_bin_prior_subset}, in this case we do not observe any bias on the cosmological parameters. However, we can appreciate a $\sim0.5\,\sigma$ shift on the $\Delta z$ parameter of the highest redshift bin 13. This is not surprising given the fact that the magnification signal becomes progressively more important at higher redshift. The contrast of these results with respect to the galaxy bias case can be understood as magnification bias is a subdominant effect compared to galaxy bias, so mis-modelling in the former is less evident than in the latter.

\subsection{Shear multiplicative bias}
In the ideal case, multiplicative bias can be calibrated, and the $m_i$ parameters can be fixed to zero. In practice, we expect a residual multiplicative bias to remain after calibration. Its magnitude can be estimated and used to set a Gaussian prior on the $m_i$ parameters. Here, we study the impact of the choice of prior on the cosmological parameters.

Note that our baseline prior width is $0.002$, which is 4 times larger than the 5$\times$$10^{-4}$ used in \cite{Canas-Herrera24}. However, we have checked that the constraints on the cosmological parameters remain indistinguishable. This implies that the size of the prior on the $m_i$ parameters has an insignificant impact on the constraining power.

\begin{figure}
    \centering
    \includegraphics[width=\linewidth]{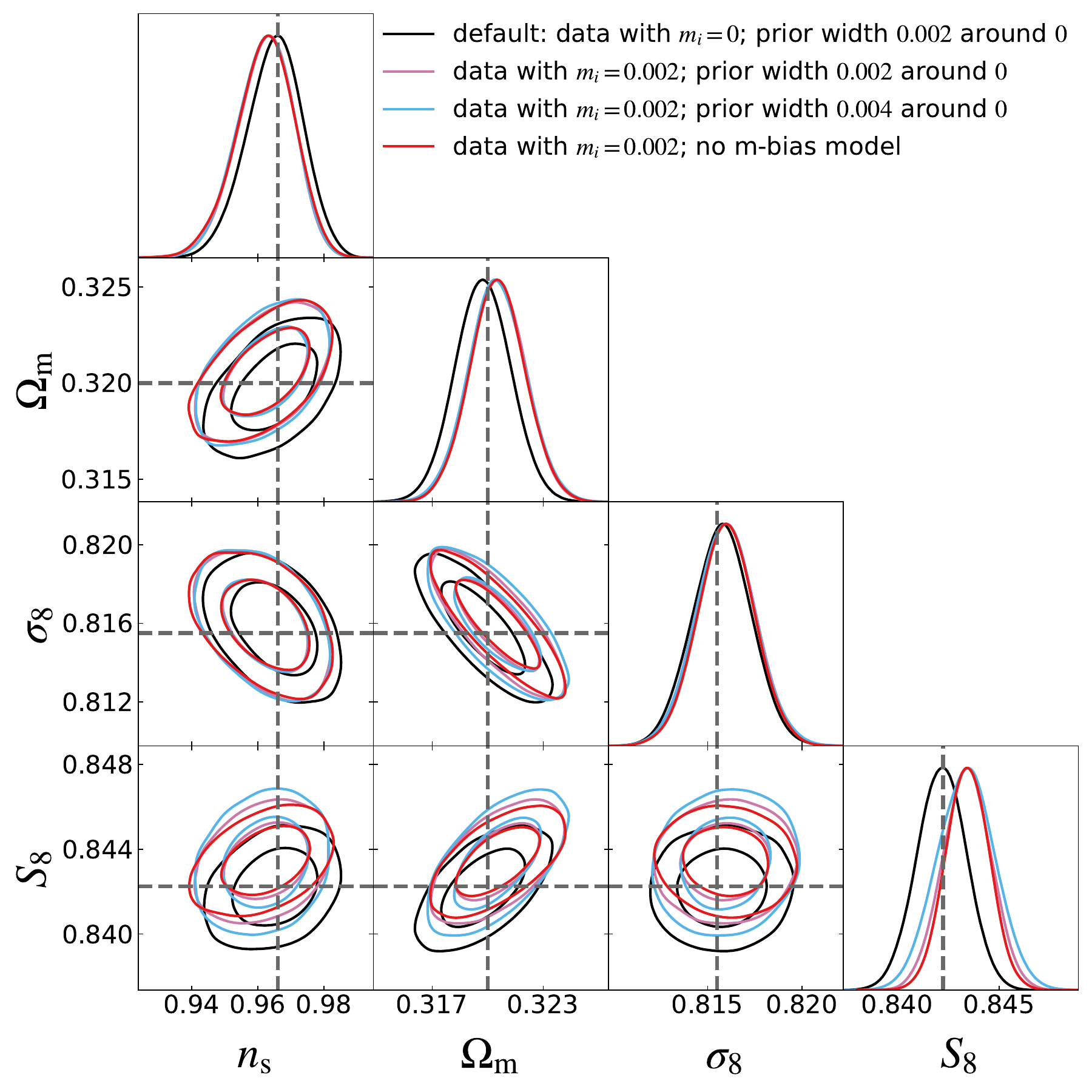}
    \caption{Cosmological parameter constraints in the baseline case (black), and analysing a data vector with $m_i=0.002$ with a prior centred at 0 with width 0.002 (purple), with a prior centred at 0 with width 0.004 (blue), and without modelling multiplicative bias (red).}
    \label{fig:shift_m_bias_broader_m_bias_prior_subset}
\end{figure}

The first case we examine is the impact of centring the prior at zero when the true value is $1\,\sigma$ away from the centre, where $\sigma$ here denotes the width of the prior. The results are shown in \cref{fig:shift_m_bias_broader_m_bias_prior_subset}. We observe a moderate bias of $0.37 \,\sigma$ and $0.72\, \sigma$ on $\Om$ and $S_8$, respectively. We then study whether doubling the prior width can resolve this issue, and find very little change, with a bias of $0.35 \,\sigma$ and $0.63\, \sigma$ on $\Om$ and $S_8$, respectively. This is almost as bad as not modelling multiplicative bias at all, which is shown in red in \cref{fig:shift_m_bias_broader_m_bias_prior_subset}.

These results indicate that care must be taken to ensure that the priors on multiplicative bias are not too narrow compared to the expected amplitude of the $m_i$ parameters.

\subsection{Shift in the mean of the redshift distribution}
In our baseline setup, we model the error on the estimation of the redshift distribution by having one nuisance parameter per redshift bin that shifts the mean of the distribution while leaving the shape unchanged. The priors on these shift parameters $\Delta z_i$ are redshift dependent and scale as 2$\times10^{-3}\,(1+z_i)$, where $z_i$ is the centre of the redshift bin \citep{EuclidSkyOverview}.

\begin{figure}
    \centering
    \includegraphics[width=\linewidth]{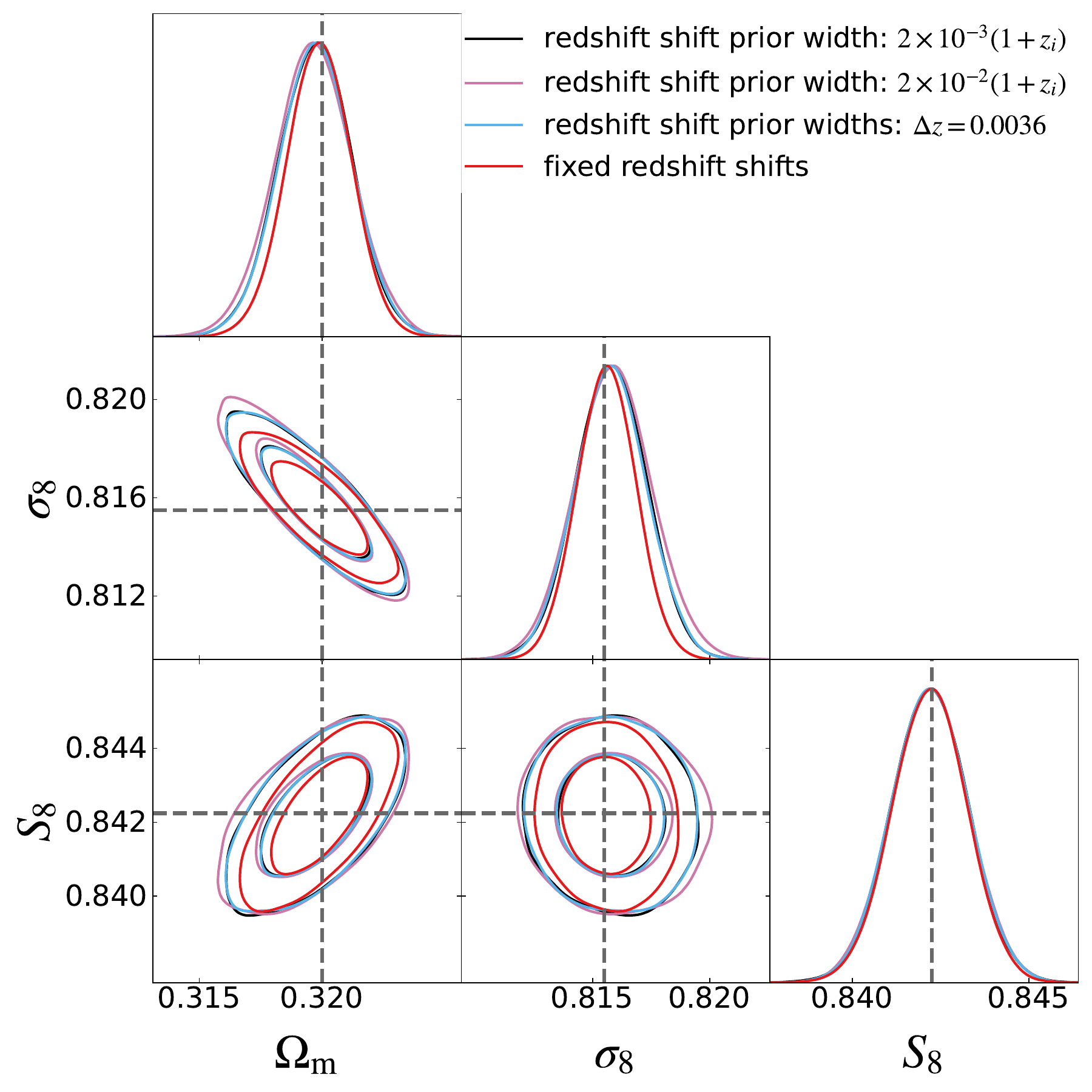}
    \caption{Cosmological parameter constraints in the baseline case (black), and using a 10 times larger prior (purple), a fully correlated prior with width $0.0036$ (blue), and fixing the redshift shifts (red).}
    \label{fig:subset_broader_photo_z_cosmo_params}
\end{figure}

To check what the impact of the size of the prior on the cosmological parameters is, we run a case where the width of the prior is ten times larger. The results are shown in \cref{fig:subset_broader_photo_z_cosmo_params} and \cref{tab:merged-constraint-shifts}. As expected, we observe a loss of constraining power of $7.32\%$ in $\Om$ and $9.81\%$ in \sotto, but their combination $S_8$ is only marginally affected, with a difference of less than $1\%$.

Although it is reasonable to expect that the errors in the photometric redshift estimation increase at higher redshift, one could opt for using fully correlated $\Delta z$ priors with a fixed width for all redshift bins, thus significantly reducing the size of the parameter space to be explored. As can be seen in \cref{fig:subset_broader_photo_z_cosmo_params}, this leads to an insignificant change in the contours. This can be understood by looking at \cref{fig:broader_photo_z}, where we show the size of the posterior compared to the prior. The $\Delta z$ parameters are well constrained by data, so changes in prior width have minimal impact on the posterior.

We also study a case where we fix all $\Delta z $ parameters to their reference value, resulting in a gain in constraining power of $11.69\%$ on $\Om$, $17.38\%$ on \sotto, and $3.97\%$ on $S_8$. This represents the maximum gain that one can achieve by perfectly calibrating the redshift distributions. This is consistent with what was recently shown in \citet{KIDSLegacy} using KiDS Legacy data, where improvements in the calibration of the redshift distributions lead to a $15\%$ gain in constraining power.
We refer the reader to Euclid Collaboration: Bertmann et al. in prep. for a more detailed study of the impact of redshift distribution uncertainties in the context of the first \Euclid Data Release.

\subsection{Purity}
\begin{figure}
    \centering
    \includegraphics[width=\linewidth]{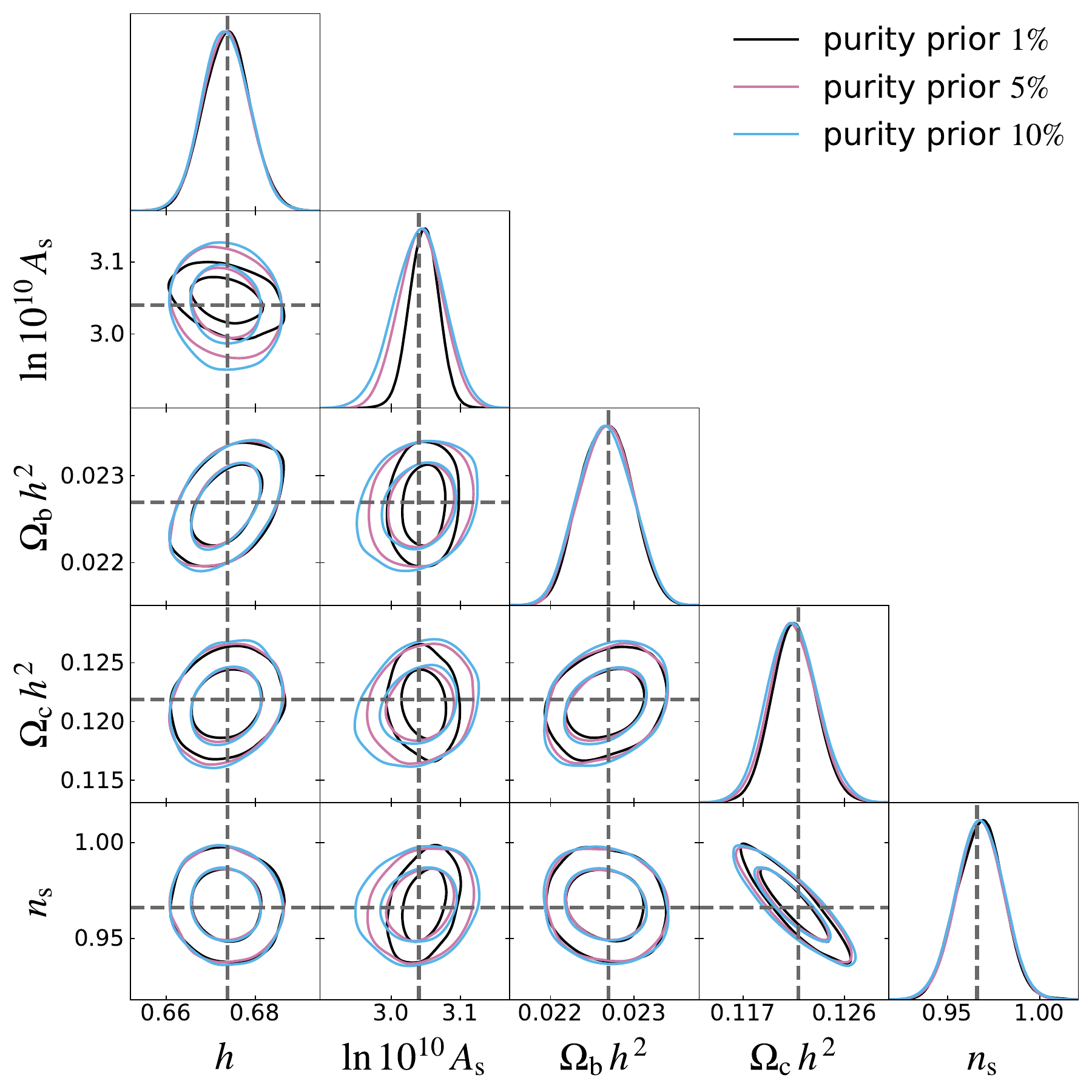}
    \caption{Cosmological parameter constraints in the baseline case (black), and using a $5\%$ prior (purple) and a $10\%$ prior (blue) on the fraction of outliers.}
    \label{fig:cosmology_GCsp}
\end{figure}

The size of the prior on the $f_{\rm out}$ parameters reflects our confidence in the estimate of the purity of the sample of $\Halpha$ galaxies. In our baseline setup, we assume that we can calibrate the fraction of outliers to within an absolute error of $1\%$, but this might not be achievable in practice. Therefore, we test the impact of having larger priors on the $f_{\rm out}$ parameters. Results are shown in Figs. \ref{fig:cosmology_GCsp} and \ref{fig:nuisance_GCsp}. As expected, the most affected parameter is $\As$, which is degenerate with the purity correction in the data vector. Having a $5\%$ ($10\%$) prior width on the outlier fraction parameters decreases the constraining power on $\As$ by $48.52\%$ ($70.23\%$) and on $\Oc h^2$ by $6.68\%$ ($13.87\%$). We also observe a loss in constraining power on the bias and EFT counter-terms parameters, as can be seen in \cref{fig:nuisance_GCsp}.

\section{Discussion and conclusions}\label{sec:discussion}
In this work, we have studied the impact of modelling and prior choices for the most important systematic effects in the 3$\times$2 pt and spectroscopic galaxy clustering 2pt analyses using synthetic data vectors that mimic the final \Euclid data release. We have identified several potentially problematic choices that warrant further study.

Among the systematic effects studied here, intrinsic alignment seems to be the most worrisome. For a signal that is expected to be small, using complex models with many parameters can lead to important biases due to prior-volume effects. At the same time, mis-modelling the signal also has an impact on the cosmological parameters, leading to biases of a sizeable fraction of the statistical uncertainty.

From our tests, we can also see that in the final \Euclid data analysis, we will be somewhat sensitive to the modelling of the redshift evolution of galaxy bias. Even if we assume a reasonable piecewise linear function of redshift, if the true underlying redshift evolution is a more complex function, we could get biases of the order of $0.5\,\sigma$ on some of the cosmological parameters. Along these lines, we should mention that, in this work, we restrict ourselves to tests of linear galaxy models and leave the small-scale modelling of the 3$\times$2pt analysis, like non-linear galaxy bias, various scale cuts as well as baryonic effects to be dealt with in the parallel work of Euclid Collaboration: Carrilho et al. in prep. 

For the parameters in which we expect to use Gaussian priors, care must be taken in choosing a width that ensures unbiased constraints while maximising the cosmological information.

For instance, we caution against using too optimistic priors on the multiplicative bias parameters, as already having a true value $1\,\sigma$ away from the centre of the prior can induce a considerable bias on the cosmological parameters.

In our setup, the redshift shifts are well constrained, so that the choice of prior width has little impact on the best-fit parameters. Enlarging the priors does however result in a loss of constraining power on the cosmological parameters. Since these parameters seem to be well constrained by the data, it is worth choosing a narrower prior width to gain constraining power. 

At the same time, the redshift shift parameters are severely impacted by mis-modelling of other systematic effects. For this reason, it is important that the redshift distributions are well calibrated. Indeed, this has been identified by the community as one of the areas that needs the most improvement. \citet{KIDSLegacy} showed the significant impact of doing so by improving their constraining power by 15\% and concurrently resolving the $S_8$ tension, which seems to have been driven by poorly modelled systematic effects.

On the spectroscopic galaxy clustering side, we have tested the impact of enlarging the prior on the fraction of outliers. As expected, this has a significant impact on the constraining power of $\As$, since the two are fully degenerate, but also on $\Oc h^2$. From this test, it is clear that a better calibration of the purity can result in significant gains in constraining power from \Euclid spectroscopic galaxy clustering alone.

This work has set the stage for more focused analyses that will consider each of these systematic effects and determine modelling and prior choices for the future \Euclid data analyses.

\begin{acknowledgements}
KT is supported by the STFC grant ST/W000903/1 and by
the European Structural and Investment Fund.
\AckEC
The authors acknowledge the contribution of the Lorentz Center (Leiden), and of the European Space Agency (ESA), where the workshop "Making \CLOE shine" and the "\CLOE workshop 2023" were held. The authors acknowledge the use of the IRIS computing facility in the UK.
\end{acknowledgements}

\bibliography{biblio,Euclid}
\begin{appendix}
\onecolumn
\section{Nuisance parameter constraints}\label{sec:appendix}
In this appendix, we show the parameter constraints of the relevant nuisance parameters for the different cases explored in Sect.~\ref{sec:results}.
\begin{figure*}[!h]
    \centering
    \includegraphics[width=1\linewidth]{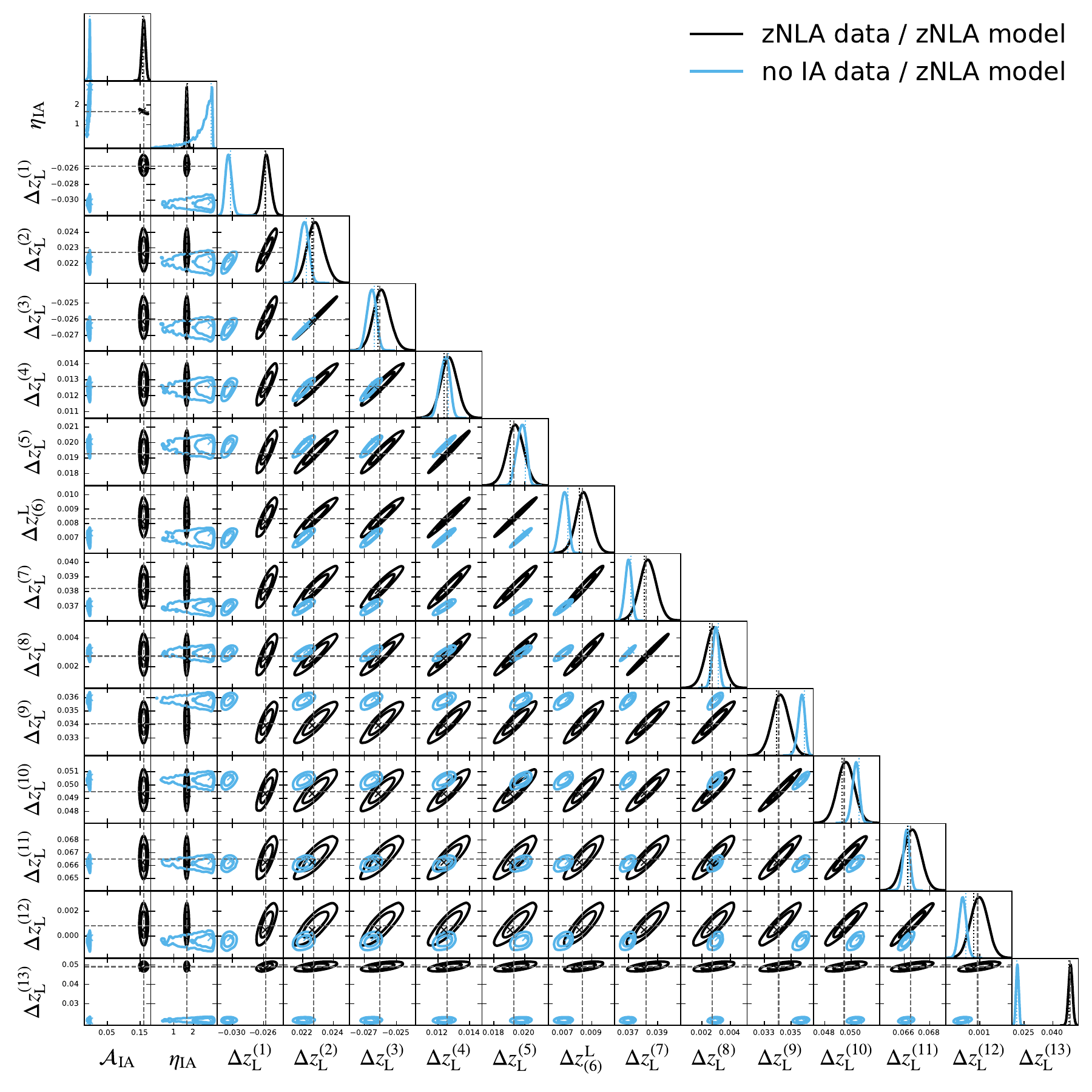}
    \caption{Nuisance parameter constraints in the baseline case (black), and using the zNLA intrinsic alignment model for a data vector with zero intrinsic alignment signal (blue). The MAP value is shown in the respective colours for the 2D (crosses) and 1D (dotted lines) projected posterior panels.}
    \label{fig:noIA_zNLA_nuis}
\end{figure*}
\begin{figure*}
    \centering
    \includegraphics[width=1\linewidth]{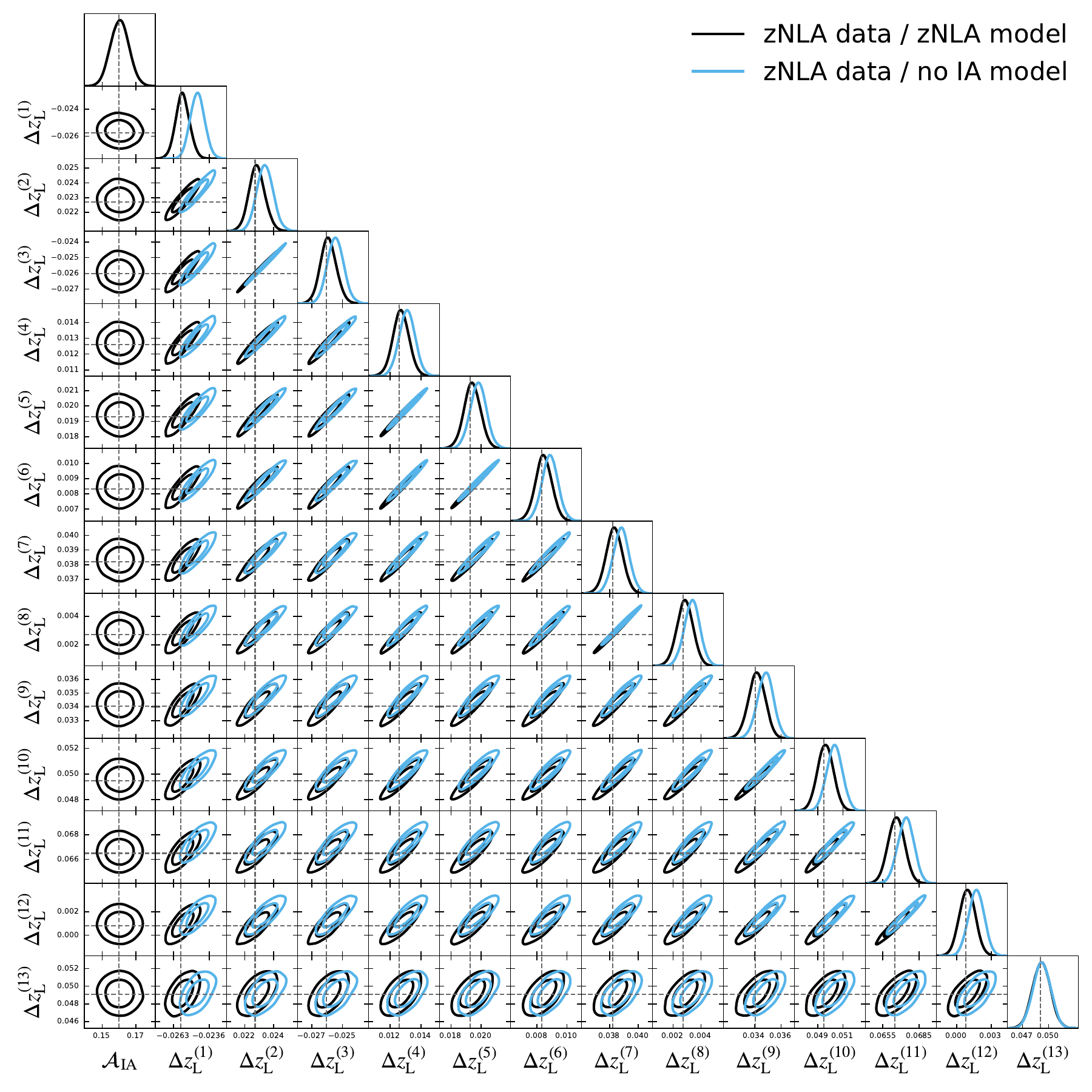}
    \caption{Nuisance parameter constraints in the baseline case (black), and analysing a data vector with zNLA signal with no intrinsic alignment model (blue).}
    \label{fig:IA_mismodelling_nuis}
\end{figure*}

\begin{figure*}
    \centering
    \includegraphics[width=1\linewidth]{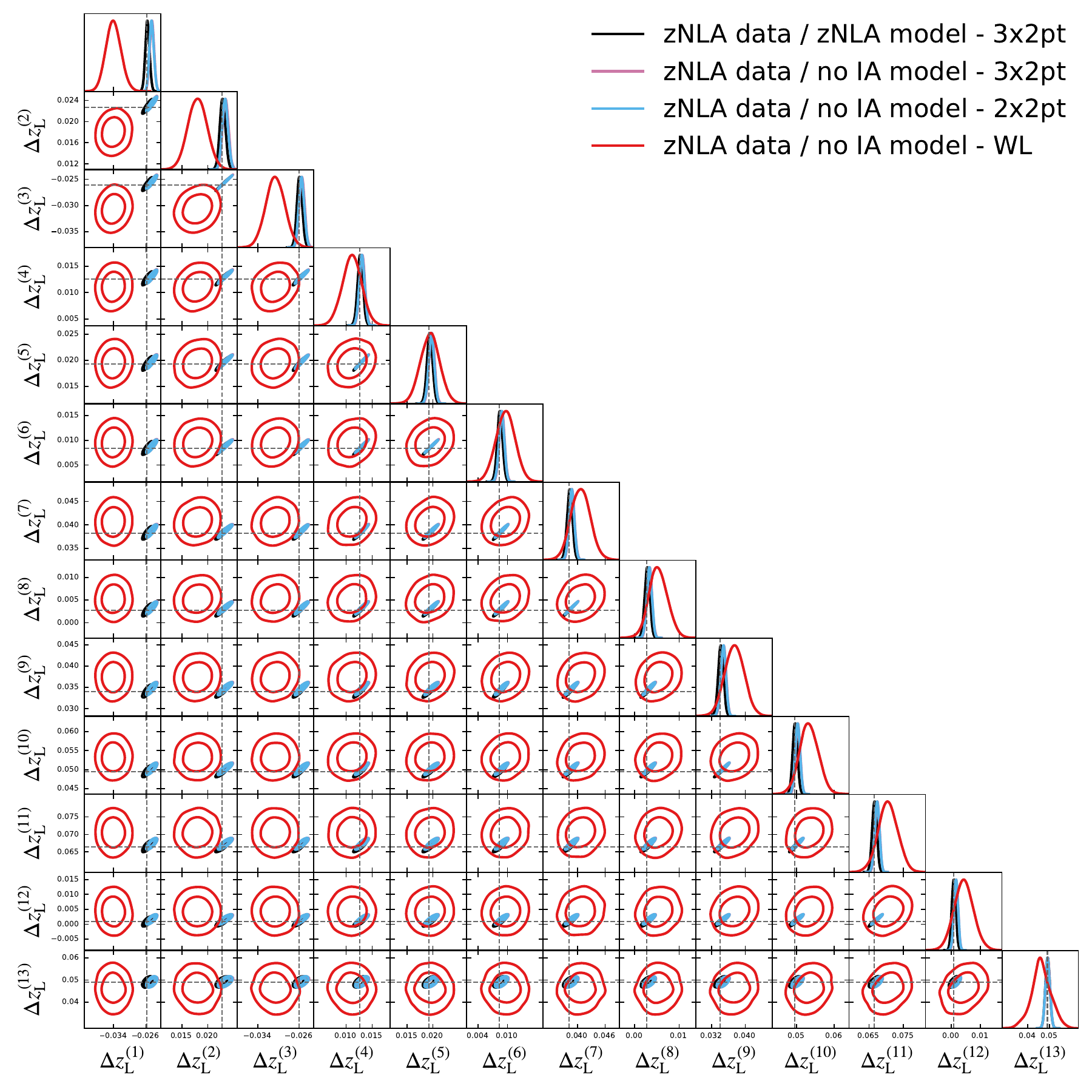}
    \caption{Nuisance parameter constraints in the baseline case (black), and analysing a data vector with zNLA signal with no intrinsic alignment model using the $3 \times 2$ pt combination (purple), $2 \times 2$ pt (blue) and WL only (red). \textit{Note:} The purple and blue contours are overlapping due to the large range of the axes, but do not coincide.}
    \label{fig:IA_mismodeling_split_nuis}
\end{figure*}

\begin{figure*}
    \centering
    \includegraphics[width=\linewidth]{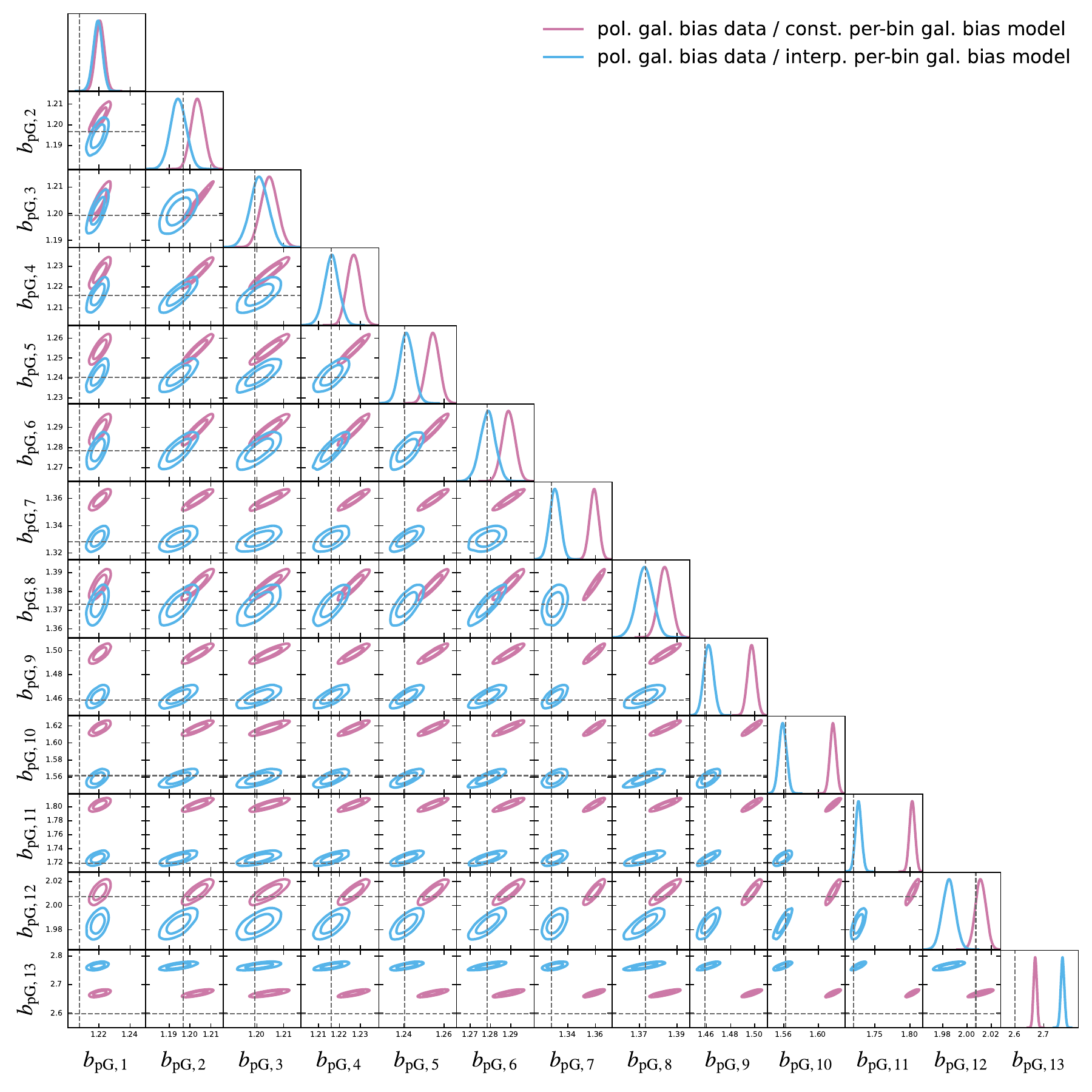}
    \caption{Constraints on the per-bin galaxy bias parameters when analysing the baseline data vector with a constant per-bin galaxy bias model (purple) or with a linear function of redshift within each bin (blue).}
    \label{fig:poly_gal_bias_data_per_bin_model_b_params}
\end{figure*}

\begin{figure*}
    \centering
    \includegraphics[width=\linewidth]{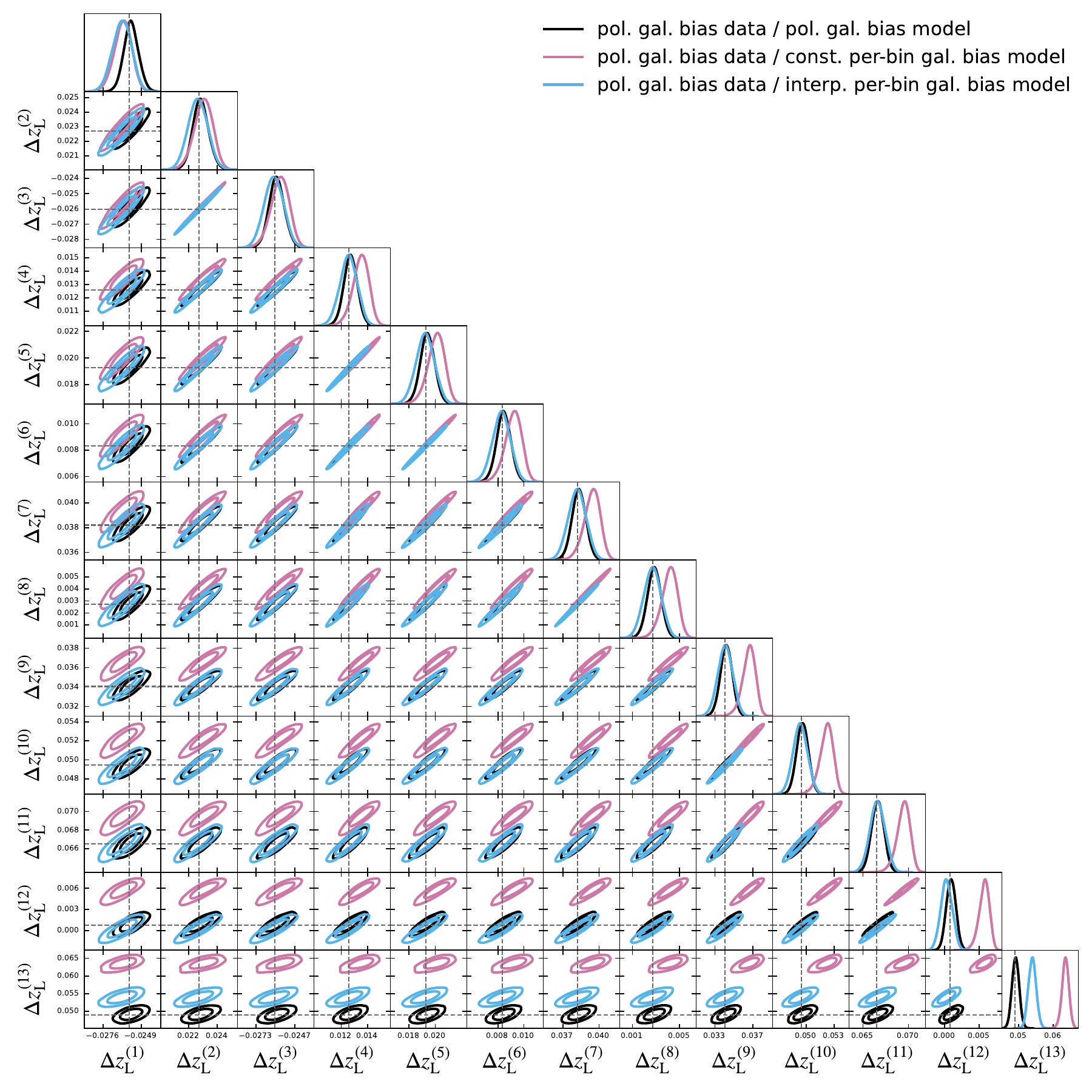}
    \caption{Constraints on the redshift shift parameters in the baseline case (black), and analysing the baseline data vector with a constant per-bin galaxy bias model (purple) or with a linear function of redshift within each bin (blue).}    \label{fig:poly_gal_bias_data_per_bin_model_b_params_effect_on_photoz}
\end{figure*}

\begin{figure*}
    \centering
    \includegraphics[width=\linewidth]{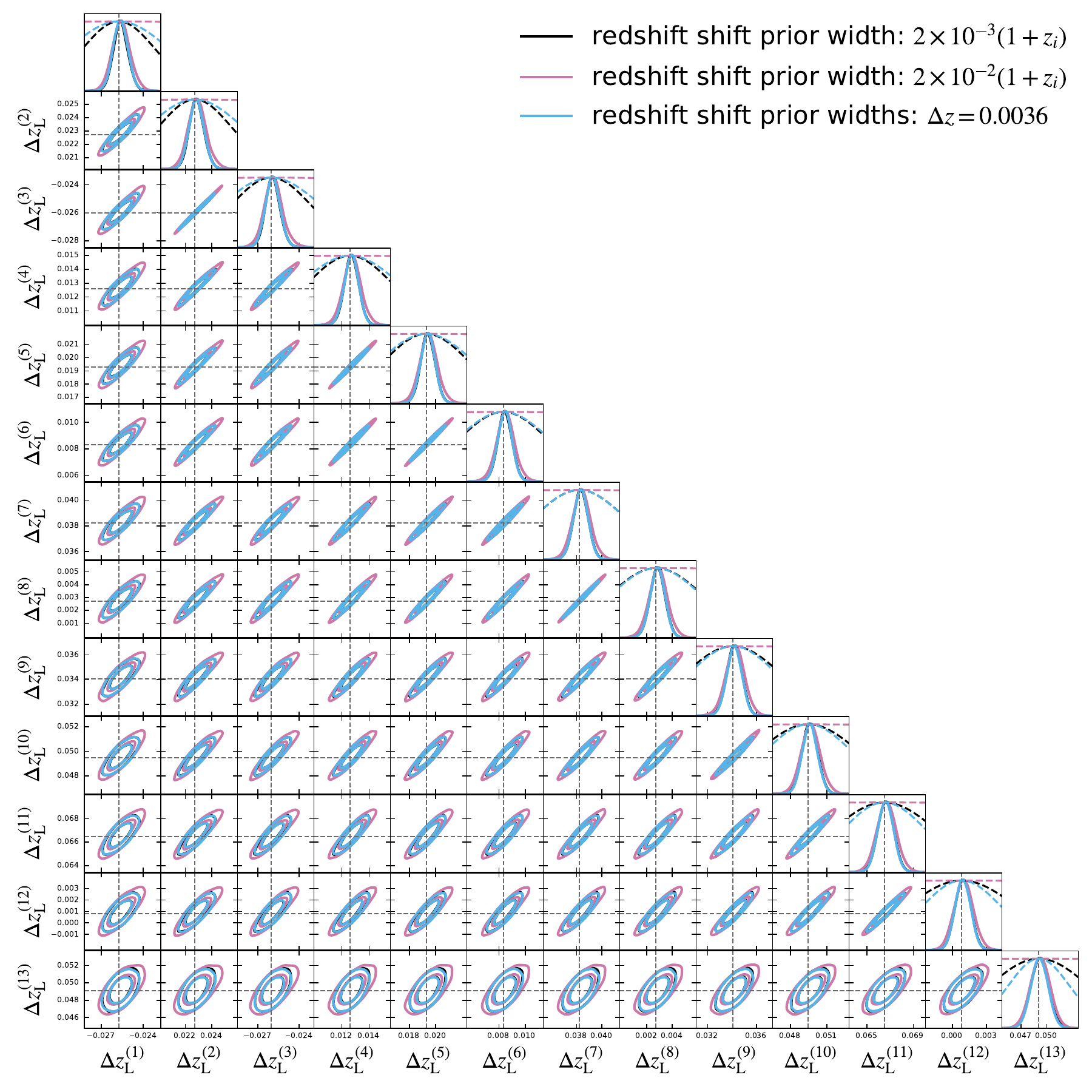}
    \caption{Redshift shift parameter constraints in the baseline case (black), and using a 10 times larger prior (purple) and a fully correlated prior with width $0.0036$ (blue). Dashed lines show the prior distribution.}
    \label{fig:broader_photo_z}
\end{figure*}

\begin{figure*}
    \centering
    \includegraphics[width=\linewidth]{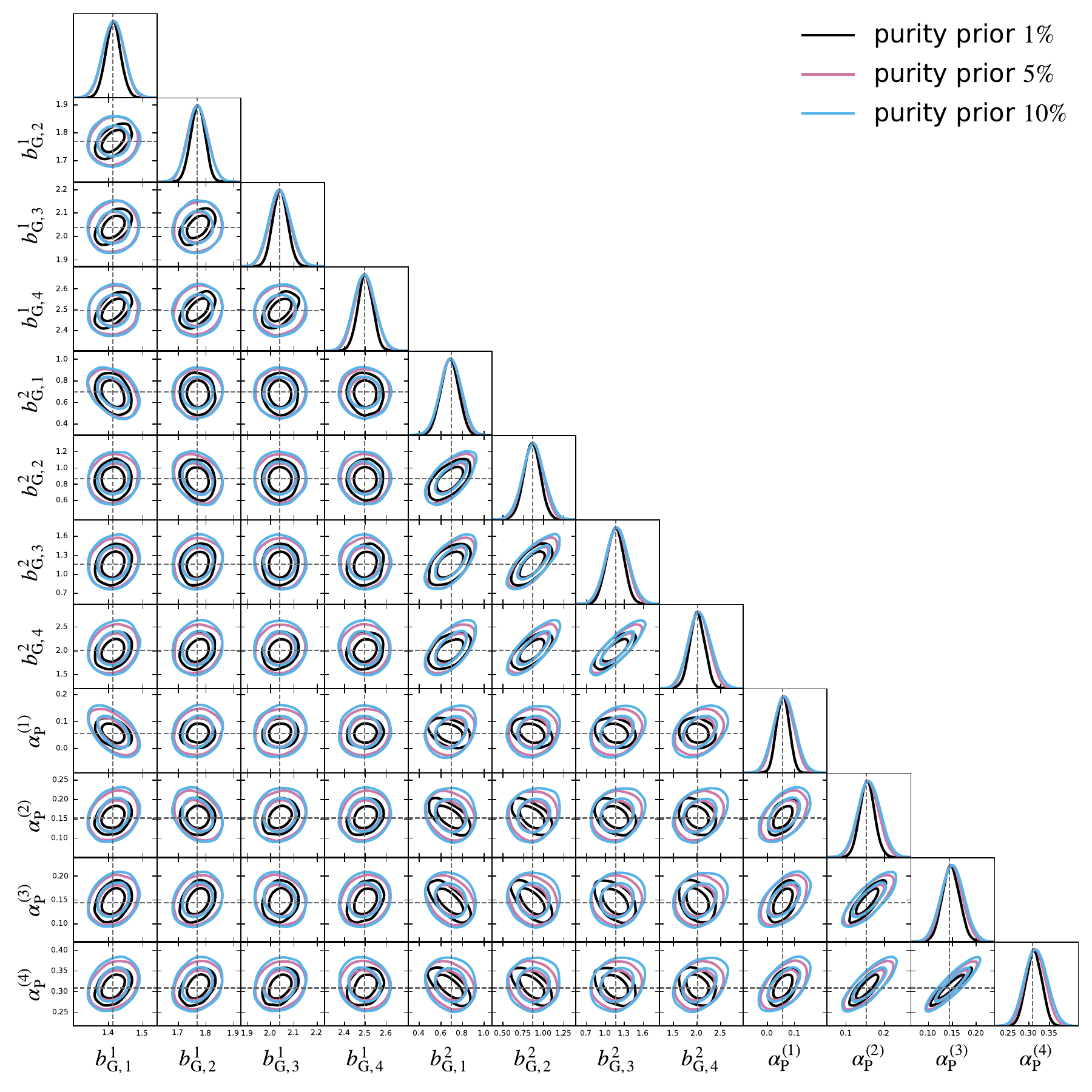}
    \caption{EFT nuisance parameter constraints in the baseline case (black), and using a 5 times larger prior (purple) and a 10 times larger prior (blue) on the fraction of outliers.}
    \label{fig:nuisance_GCsp}
\end{figure*}

\begin{figure*}
    \centering
    \includegraphics[width=\linewidth]{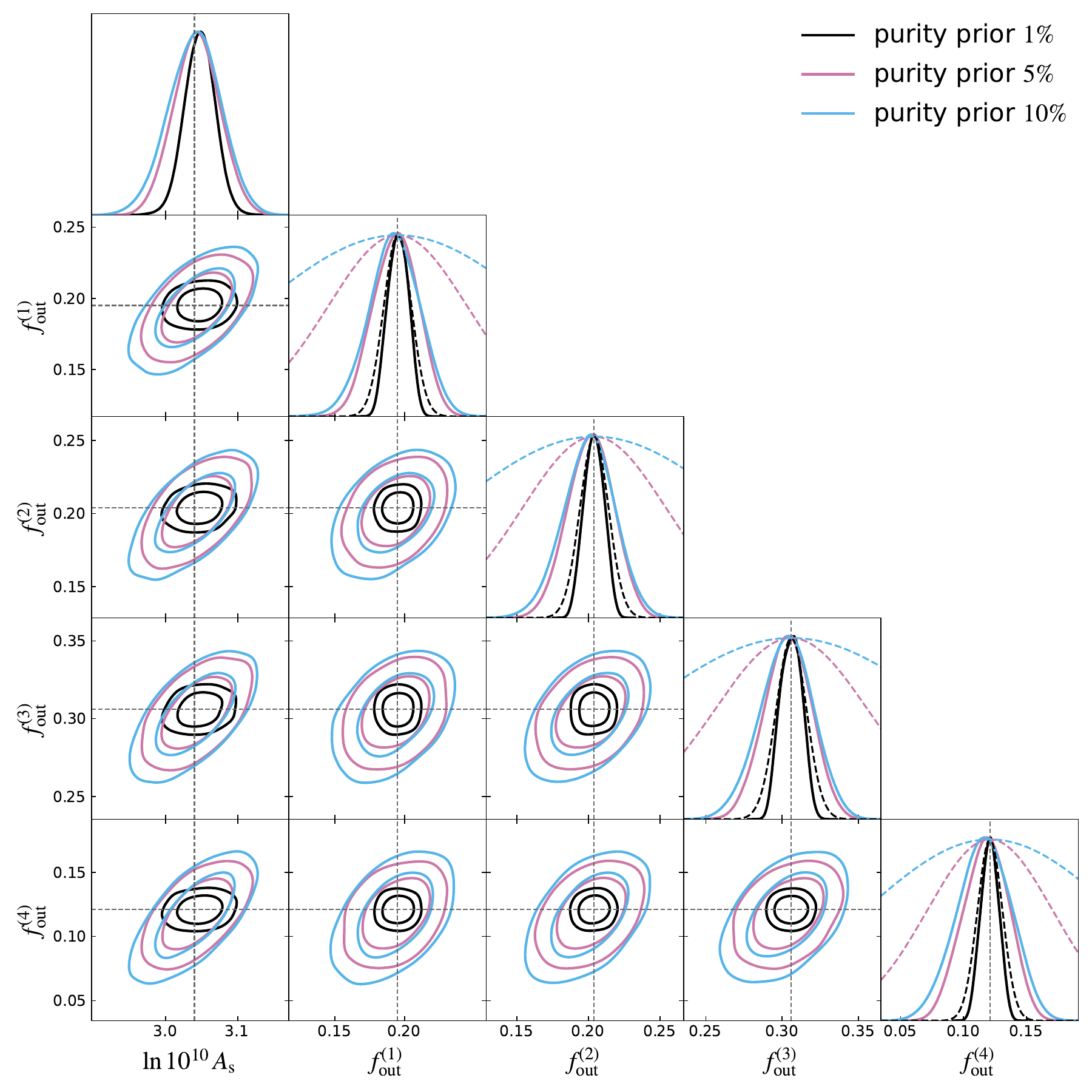}
    \caption{$f_{\rm out}$ parameter constraints versus $\lnAs$ in the baseline case (black), and using a $5\%$ prior (purple) and a $10\%$ prior (blue) on the fraction of outliers. Dashed lines show the prior distribution.}
    \label{fig:extra_GCsp}
\end{figure*}
\end{appendix}
\end{document}